\documentclass[a4paper,runningheads]{llncs}

\usepackage{amsmath}
\usepackage{amssymb}
\usepackage{enumitem}
\usepackage{xspace}
\usepackage{colortbl,xcolor}
\usepackage{comment}
\usepackage{xifthen}
\usepackage{tikz}
\usepackage{tkz-graph}

\newcommand{\repeatthanks}{\textsuperscript{\thefootnote}}

\newcommand{\Abelard}{Abelard\xspace}
\newcommand{\Ab}{\ensuremath{\forall}\xspace}
\newcommand{\Acc}{\mathcal{A}}
\newcommand{\AtomicGames}{\AtG}

\newcommand{\comp}{\mathop{;}}
\newcommand{\cross}{{\mathop{\times}}}
\newcommand{\diam}[1]{\langle {#1} \rangle}
\newcommand{\G}{\mathbb{G}} 
\newcommand{\Heloise}{Eloise\xspace}
\newcommand{\He}{\ensuremath{\exists}\xspace}

\newcommand{\Lit}{\mathsf{Lit}}
\newcommand{\Mod}{\mathsf{Mod}}
\newcommand{\Latt}{\mathsf{Latt}}
\newcommand{\Lbl}[1]{\ifthenelse{\isempty{#1}{}}
  {\mathsf{Lb}}{\mathsf{Lb}_{#1}}
} 
\newcommand{\arity}{\mathsf{arity}}
\newcommand{\M}{\mathbb{S}}

\newcommand{\Prop}{\mathsf{Prop}}
\newcommand{\range}[1]{\mathrm{ran}({#1})}

\newcommand{\reinit}[2]{{#1}@{#2}}

\newcommand{\Fix}{\mathrm{Fix}}

\newcommand{\AtG}{\mathsf{Gam}}
\newcommand{\Form}{\mathcal{F}}
\newcommand{\FormDNNF}{\Form_\mathrm{DNNF}}
\newcommand{\GT}{\mathcal{G}}
\newcommand{\GtDNNF}{\mathcal{G}_\mathrm{DNNF}}

\renewcommand{\phi}{\varphi}
\newcommand{\sse}{\subseteq}

\renewcommand{\iff}{\quad \mbox{iff} \quad}

\newcommand{\abel}{\forall}
\newcommand{\bbB}{\mathbb{B}}
\newcommand{\game}{\mathcal{F}}
\newcommand{\can}{{\tiny \mathrm{can}}}

\newcommand{\inddef}{\mathrel{::=}}
\newcommand{\isdef}{\mathrel{:=}}
\newcommand{\bang}[1]{{#1}!}
\newcommand{\iter}[1]{{#1}^*}
\newcommand{\diter}[1]{{#1}^\cross}
\newcommand{\subf}{\triangleleft}

\newcommand{\Om}{\Omega}
\newcommand{\card}[1]{\#{#1}}
\newcommand{\Pow}{\mathcal{P}}
\newcommand{\Mon}{\mathcal{M}}
\newcommand{\MFr}{\mathrm{MF}}
\newcommand{\kcomp}{\mathrel{;}}
\newcommand{\kcross}[1]{{#1}^\cross}
\newcommand{\kiter}[1]{{#1}^*}
\newcommand{\Demon}[1]{\mathsf{D}_#1}
\newcommand{\Angel}[1]{\mathsf{A}_#1}
\newcommand{\LFP}{\mathrm{LFP}}
\newcommand{\GFP}{\mathrm{GFP}}
\newcommand{\lsem}{[\![}
\newcommand{\rsem}{]\!]}
\newcommand{\sem}[1]{\lsem #1 \rsem}
\newcommand{\nsem}[1]{\langle \!|{#1} |\! \rangle}
\newcommand{\Val}{\Upsilon}
\newcommand{\Ev}{\mathcal{E}}
\newcommand{\Win}{\mathrm{Win}}
\newcommand{\E}{\exists}

\newcommand{\Suc}{\mathrm{Suc}}
\newcommand{\play}{\Pi}
\newcommand{\dnnf}[1]{\mathrm{nf}(#1)}
\newcommand{\dtest}[1]{{#1}!}

\newcommand{\GLGraphs}{\mathbf{GG}} 

\newcommand{\matchingConnective}{\textbf{(parity)}\xspace}
\newcommand\brexit{\textbf{(head)}\xspace}
\newcommand\leavingCycles{\textbf{(leave)}\xspace}
\newcommand\noRemainPathToHigher{\textbf{(remain)}\xspace}
\newcommand{\exitPath}{\textbf{(exit)}\xspace}

\newcommand{\ra}{\rightarrow}
\newcommand{\Ra}{\Rightarrow}

\newcommand{\lra}{\leftrightarrow}
\newcommand{\pfun}{\rightharpoonup}

\newcommand{\sqsse}{\sqsubseteq}

\newcommand{\blue}[1]{#1}
\newcommand{\red}[1]{#1}

\newcommand{\weg}[1]{}

\newtheorem{claim2}{\sc Claim}
\newenvironment{yclaim}{\begin{claim2}\rm}{\end{claim2}\rm}

\newenvironment{proofof}[1]{\begin{trivlist}\item[\hskip\labelsep{\bf
Proof~of~{#1}.\ }]}{\hspace*{\fill} {\qed}\end{trivlist}}

\usepackage{ifthen}

\newtheorem{definition_th}{Definition}[section]  

\newtheorem{algo_th}[definition_th]{Algorithm}

\newenvironment{theorem_for}[2][\empty]{\bigskip\noindent{\bf
    Theorem~\ref{#2}}\ifthenelse{\equal{#1}{\empty}}{{\bf.}}{ {\bf (#1).}}\it}{\vspace{0.5cm}}

\newenvironment{proposition_for}[2][\empty]{\bigskip\noindent{\bf
    Proposition~\ref{#2}}\ifthenelse{\equal{#1}{\empty}}{{\bf.}}{ {\bf
      (#1).}}\it}{\vspace{0.5cm}}
\newenvironment{lemma_for}[2][\empty]{\bigskip\noindent{\bf
    Lemma~\ref{#2}}\ifthenelse{\equal{#1}{\empty}}{{\bf.}}{ {\bf (#1).}}\it}{\vspace{0.5cm}}

\newif\ifWithProofs
\WithProofstrue   
\WithProofsfalse   

\title{Parity Games and Automata for Game Logic}

\author{
  Helle Hvid Hansen\inst{1} \and
  Clemens Kupke \thanks{Supported by EPSRC grant EP/N015843/1.} \inst{2} \and
  Johannes Marti \repeatthanks \inst{2} \and
  Yde Venema\inst{3}
}
\institute{
  Delft University of Technology, Delft, The Netherlands \and
  University of Strathclyde, Glasgow, United Kingdom \and
  University of Amsterdam, Amsterdam, The Netherlands
}

\begin{document}

\maketitle

\begin{abstract}
  Parikh's game logic is a PDL-like fixpoint logic interpreted on
monotone neighbourhood frames that represent the strategic power of
players in determined two-player games. Game logic translates into a
fragment of the monotone $\mu$-calculus, which in turn is expressively equivalent to monotone modal automata.
Parity games and automata are important tools for dealing with the
combinatorial complexity of nested fixpoints in modal fixpoint logics,
such as the modal $\mu$-calculus. In this paper, we (1) discuss the
semantics a of game logic over neighbourhood structures in terms of
parity games, and (2) use these games to obtain an automata-theoretic
characterisation of the fragment of the monotone $\mu$-calculus that
corresponds to game logic.
Our proof makes extensive use of structures that we call syntax graphs
that combine the ease-of-use of syntax trees of formulas with the
flexibility and succinctness of automata. They are essentially a
graph-based view of the alternating tree automata that were introduced
by Wilke in the study of modal $\mu$-calculus.
\end{abstract}

\section{Introduction}

Game logic was introduced by Parikh \cite{Parikh85} as a modal logic for
reasoning about strategic power in determined 2-player games, and
it can be seen as a generalisation of PDL \cite{HKT:Dynamic-Logic}
both in terms of syntax and semantics.
On the syntax side,
game logic is a multi-modal language in which modalities are labelled by games,
which in turn are built from atomic games, the PDL program constructs together
with the operation \emph{dual} which switches the role of the players.
A modal formula $\diam{\alpha}\phi$ should be read as \emph{``player 1 has a strategy in the game $\alpha$ to achieve an outcome that satisfies the formula $\phi$''}.
On the semantic side,
one goes from PDL to game logic by moving from Kripke frames to
monotone neighbourhood frames.
A game perspective on this generalisation is that nondeterministic programs (i.e., relations) are 1-player games in which the player chooses his move from a set of successors, and monotone neighbourhood frames are 2-player games where player 1 first chooses a neighbourhood $U$, and then player 2 chooses an element in $U$.
The shift from Kripke frames to monotone neighbourhood frames also means that
we go from normal modal logic to monotone modal logic.
Just as PDL (and other fixpoint logics such as LTL and CTL$^*$)
can be viewed as a fragment of the modal $\mu$-calculus \cite{koze:resu83,brad:moda06},
game logic can be naturally viewed as a fragment of the
\emph{monotone $\mu$-calculus} \cite{Pau:phd},
which is monotone (multi-) modal logic with explicit fixpoint operators.
A notable difference is that PDL, LTL and CTL$^*$ are all contained in level 1 or 2 of the alternation hierarchy whereas game logic, due to the combination of dual and iteration, spans all levels of the alternation hierarchy \cite{Berwanger:GL-parity}. This high level of expressiveness could be an explanation for why a completeness proof for game logic is still missing.

 In this paper we contribute to the theory of game logic. We discuss the
semantics of game logic over neighbourhood structures using parity games
and then use these games to characterise a class of automata that is
exactly as expressive as formulas in game logic. Parity games
are an intuitive way of dealing with the nesting of least and greatest
fixpoint operators, and together with automata they play a fundamental
role in the theory of fixpoint logics \cite{grae:auto02}.
  For instance, parity games and automata have been used in proving
complexity results for the modal $\mu$-calculus
\cite{emer:mumc01,emer:comp88} and also Walukiewicz' completeness
result~\cite{walu:comp93} is proved by automata-theoretic means.
  Some of these results have been extended to the setting of coalgebraic
fixpoint logic~\cite{enqv:comp17}. In particular, they are applicable to
the monotone $\mu$-calculus. Since monotone modal $\mu$-calculus is
expressively equivalent to a naturally defined class of
\emph{(unguarded) monotone modal automata} \cite{font:auto10}, it is of
interest to find out which subclass of these automata corresponds to
game logic. The main result in our paper is a characterisation of a
class of unguarded monotone modal automata that effectively corresponds
to game logic, in the sense that there are effective translations in
both directions. This result can be seen as the game logic analogue of
the characterisation of PDL in automata-theoretic
terms~\cite{carr:pdl2014}. The case of game logic, however, is more
involved because composition of games does not distribute from the left
over choice as is the case for the programs in PDL. This is related to
the fact that in the relational semantics of PDL, diamonds distribute
over disjunctions; this property, which is heavily exploited in the
mentioned results on PDL, does not apply to the diamonds of game logic.
Finally, note that our characterisation can also be seen as an
automata-theoretic counterpart to the results in
\cite[sec.~3.3]{carr:fofl2015} that characterise a fragment of the
$\mu$-calculus that is expressively equivalent to game logic interpreted
over Kripke frames.

   Our characterisation goes via a class of structures that we call \emph{syntax graphs}.
   Syntax graphs combine the ease-of-use of syntax trees of formulas with the flexibility
   and succinctness of automata.
   They are essentially the same as Wilke's alternating tree automata (ATAs) \cite{Wilke01:ATA} except they are described in terms of their transition graphs, and they run on monotone neighbourhood models rather than Kripke models.
   Unguarded monotone modal automata can, in turn, be viewed as
   Wilke's ATAs with complex transition condition \cite{Wilke01:ATA} (again with a semantics over monotone neighbourhood models). As noted in \cite{Wilke01:ATA,Kirsten02:ATA} an ATA with complex transition conditions can be effectively translated into an equivalent ATA, and this construction is easily seen to work also for monotone semantics.
   Concretely, our characterisation consists of a number of conditions that define
   a subclass $\GLGraphs$ of syntax graphs
   that correspond to game logic formulas. We call these \emph{game logic graphs}.
   A game automaton is then a monotone modal automaton whose corresponding syntax graph (i.e. ATA) is in $\GLGraphs$.
   The translation from formulas to game logic graphs is an inductive construction similar
   to the construction of a nondeterministic automaton from a regular expression.
   Conversely,
   the defining conditions on game logic graphs allow us
   to decompose a game logic graph into components that correspond to formulas.

   The rest of the paper is structured as follows. In Section~\ref{game
logic} we recall the syntax and neighbourhood semantics of game logic
and describe a normal form that is needed for our results. In
Section~\ref{sec:games} we introduce the game semantics for game logic
and prove it to be equivalent to the neighbourhood semantics. In
Section~\ref{sec:graphs} we discuss syntax graphs and their game
semantics. In Section~\ref{sec:fragment} we define game logic graphs and
prove them to be expressively equivalent to formulas in game logic.
Due to space constraints, proofs are provided in the appendix.

\section{Game Logic}
\label{game logic}


Most definitions and results in this section are from
\cite{Parikh85,PaulyParikh:GL}.
The syntax of game logic is based on the syntax of propositional modal logic 
with the additional feature that modal operators are labelled with terms that denote games. Since we have ``test games'' of the form $\varphi ?$, the definition of the syntax is a simultaneous recursion on the structure of formulas and games. 

\begin{definition}
	Throughout the paper we fix a countable set $\Prop$ of atomic propositions (proposition letters) and a set $\AtG$ of atomic games.
	The sets $\Form$ of formulas and $\GT$ of game terms of game logic are defined recursively as follows:
	\begin{eqnarray*} 
	\Form \ni \varphi & \inddef & p \in \Prop  \mid \neg \varphi \mid \varphi \land \varphi \mid \varphi \lor \varphi \mid \diam{\alpha} \varphi, \quad \text{where } \alpha \in \GT  \\
	\GT \ni \alpha & \inddef & g \in \AtG \mid \alpha^d \mid \alpha \cup \alpha \mid  \alpha \cap \alpha \mid \alpha;\alpha \mid \iter{\alpha} \mid \diter{\alpha} \mid {\varphi}? \mid \bang{\varphi},\quad \text{where } \varphi \in \Form
	\end{eqnarray*}
        We use the standard definitions of $\ra$ and $\lra$, and note that $\top$ can be defined as $p \lor \lnot p$ for any $p \in \Prop$. 
	In the following we denote formulas by $\varphi, \psi,\dots$ and game terms with $\alpha,\beta,\rho,\dots$.
	We use the letter $\chi$ to denote arbitrary terms that could either be a formula or a game term.
\end{definition}

The formulas of game logic express strategic power in 2-player determined, zero-sum games. A formula $\diam{\alpha}\phi$ says that player 1 has a strategy in the game $\alpha$ to ensure that the outcome of the game satisfies $\phi$.
\blue{The assumption that the games are determined and zero-sum means that in a given game $\alpha$, player 2 has a strategy to achieve $\phi$ iff player 1 does not have a strategy to achieve $\lnot\phi$. Hence the formula $\lnot\diam{\alpha}\lnot\phi$, usually written as $[\alpha]\phi$, says that player 2 has a strategy in $\alpha$ to ensure an outcome that satisfies $\phi$. For technical reasons we do not include boxes as primitive operators.}

It will be convenient to refer to player 1 as Angel and player 2 as
Demon. The game operations can then be explained as follows. The
composition $\alpha \comp \beta$ is the game consisting of playing
$\alpha$ followed by $\beta$. The angelic choice $\alpha \cup \beta$
(resp.\ demonic choice $\alpha \cap \beta$) is the game in which Angel
(resp.\ Demon) chooses whether to play $\alpha$ or $\beta$. The angelic
iteration $\iter{\alpha}$ is the game in which $\alpha$ is played 0 or
more times, and after each time, Angel chooses whether to stop or play
again, but she must stop after some finite number of iterations. The
demonic iteration $\diter{\alpha}$ is the iterated game in which Demon
chooses when to stop, and he may choose to play forever. \blue{The
formula $\diam{\iter{\alpha}}\phi$ thus says that Angel has a strategy
to reach a $\phi$-state by playing $\alpha$ some finite number of rounds
(where her strategy may depend on what Demon did in previous rounds, so
that in particular, the number of rounds needed to reach $\phi$ is not
determined at the start of the game). The formula
$\diam{\diter{\alpha}}\phi$ says that Angel has a strategy for
maintaining $\phi$ indefinitely when playing $\alpha$ repeatedly.}
Finally, the dual game $\alpha^d$ is the same as $\alpha$ but with the
roles of the two players reversed, i.e., \blue{Angel has a strategy to
achieve $\phi$ in $\alpha^d$ iff Demon has a strategy to achieve $\phi$
in $\alpha$, and vice versa.}

In \cite{Parikh85,PaulyParikh:GL}, the language of game logic only contained the game operations $\comp\, , \cup, ^*, ^d$, and the demonic operations were defined as
$\alpha \cap \beta = (\alpha^d \cup \beta^d)^d$ and
$\diter{\alpha} = (\iter{(\alpha^d)})^d$. We take the demonic operations as primitives, since later we want to reduce formulas to dual and negation normal form.

The formal semantics of game logic is given by representing games as monotone neighbourhood frames. \blue{These are well known semantic structures in modal logic \cite{Chellas,Han03:MonML}.} 

\begin{definition}
  Let $S$ be a set. We denote by $\Mon(S)$ the set of up-closed subsets of $\Pow(S)$, i.e.,
  $\Mon(S) = \{ N \sse \Pow(S) \mid \forall U,U': U \in N, U \sse U' \Ra U' \in N\}$.
  A \emph{monotone neighbourhood frame on $S$} is a function $f \colon S \to \Mon(S)$.
  We denote by $\MFr(S)$ the set of all monotone neighbourhood frames on $S$.
\end{definition}

\blue{For $f \in \MFr(S)$ and $s \in S$, the subsets $U$ in $f(s)$ are called the neighbourhoods of $s$. We point out that such neighbourhoods are not necessarily neighbourhoods in the topological sense. In particular, we do not require that a state $s$ is an element of all its neighbourhoods. In our setting, the neighbourhoods will be the subsets that Angel can force in the game represented by $f$.}

We note that $(\Mon(S), \sse)$ is a complete partial order with associated join and meet given by union and intersection of neighbourhood collections. This CPO structure lifts pointwise to a CPO $(\MFr(S),\sqsse)$ in which we also denote join and meet by $\cup$ and $\cap$.


In analogue with how the PDL program operations are interpreted in relation algebra, we interpret game operations via algebraic structure on $\MFr(S)$.\footnote{It is well-known that $\Mon$ is a monad, \cite{HK:FICS15}. Readers who are familiar with monads will recognise that unit and composition correspond to the unit and Kleisli composition.}

\begin{definition}[Game operations]
  Let $f, f_1,f_2 \in \MFr(S)$ be monotone neighbourhood frames.
  We define
  \begin{itemize}
  \item the unit frame $\eta_S$ by:\quad
    $U \in \eta_S(s) \mbox{ iff } s \in U \; \qquad \mbox{ for } s \in S \mbox{ and } U \subseteq S.$ 
  \item the composition $f_1 \kcomp f_2$ by:
    \[U \in (f_1 \kcomp f_2)(s) \mbox{ iff } \{s' \in S \mid U \in f_2(s')\} \in f_1(s) \qquad \mbox{ for } s \in S \mbox{ and } U \subseteq S.\]
  \item the Angelic choice and Demonic choice between $f_1$ and $f_2$ by:
    \[
    (f \cup g)(s) = f(s) \cup g(s)
\qquad
    (f \cap g)(s) = f(s) \cap g(s), \quad \text{ for } s \in S.
\]
  \item the dual $f^d$ by: \quad
    $U \in f^d(s) \iff S\setminus U \notin f(s) \qquad \mbox{ for } s \in S \mbox{ and } U \subseteq S.$
  \item the angelic iteration $\kiter{f} \isdef \LFP(\Angel{f})$, 
  \item the demonic iteration $\kcross{f} \isdef \GFP(\Demon{f})$,
  \end{itemize}
  where $\LFP(\Angel{f})$ and $\GFP(\Demon{f})$ are the least and greatest fixed points of the maps
  \[
  \begin{array}{rrclcrrcl}
    \Angel{f}: & \MFr(S) & \to & \MFr(S) & \qquad & \Demon{f}: & \MFr(S) & \to & \MFr(S)  \\
    & g & \mapsto & \eta_S \cup (f \kcomp g) & \qquad\qquad & \qquad & g & \mapsto & \eta_S \cap (f \kcomp g)
  \end{array}
  \] 
  Note that for any $f \in \MFr(S)$, the map $g \mapsto f \kcomp g$ is a monotone operation on $(\MFr(S),\sqsse)$ and hence so are $\Angel{f}$ and $\Demon{f}$.
  By the Knaster-Tarski theorem, $\Angel{f}$ and $\Demon{f}$ have unique least and greatest fixed points.
\end{definition}

It is straightforward to verify that $\MFr(S)$ is closed under the above operations.
The following lemma lists a number of identities that will be useful in reasoning about game logic semantics.

\def\stateLemMFcongruence{
  For all $f,g \in \MFr(S)$, we have:\\[.5em]
  \begin{tabular}{llll}
  1. $(f^d)^d = f$ \qquad&
  4. $(f \cup g)^d = f^d \cap g^d$ \qquad &
  7. $(\iter{f})^d = \diter{(f^d)}$ \\
  2. $(f \kcomp g)^d = f^d \kcomp g^d$ \quad\qquad &
  5. $(f \cap g)^d = f^d \cup g^d$ \quad\qquad &
  8. $(\diter{f})^d = \iter{(f^d)}$ \\
  3. $(\eta_S)^d = \eta_S$ \qquad &
  6. $f \sse g \;\;\Rightarrow\;\; g^d \sse f^d$ \qquad &
  \end{tabular}
}
\begin{lemma}\label{lem:MF-congruence}
\stateLemMFcongruence
\end{lemma}

\def\proofLemMFcongruence{
\begin{proof}
  Items 1, 3, 4, 5, 6 are straightforward.
  To see why item 2 holds:
  \[\begin{array}{rclrll}
  U \in (f\kcomp g)^d(s)
  & \iff & S \setminus U \notin (f \kcomp g)(s) \\
  \text{(by mon.)} & \iff & \forall V \in f(s)\; \exists t \in V : S\setminus U \notin g(t)\\
  & \iff & \{ s' \in S \mid S\setminus U \in g(s')\} \notin f(s) \\
  & \iff &  \{ s' \in S \mid S\setminus U \notin g(s')\} \in f^d(s) \\
  & \iff &  \{ s' \in S \mid U \in g^d(s')\} \in f^d(s) \\
  & \iff & U \in (f^d \kcomp g^d)(s)
  \end{array}\]
  Item 7: First observe that
  $U \in (\iter{f})^d(s)$  iff $S\setminus U \notin \iter{f}(s) = \LFP(\Angel{f})(s)$
  iff
\begin{equation}\label{eq:congr-star-dual}
  \text{there is a } g \in \MFr(S): S\setminus U \notin g(s) \text{ and } \eta_S(s) \cup (f \kcomp g)(s) \sse g(s)
\end{equation}
  and
  $U \in \diter{(f^d)}(s) = \GFP(\Demon{{f^d}})$
  iff
\begin{equation}\label{eq:congr-dual-cross}
  \text{there is a } h \in \MFr(S): U \in h(s) \text{ and } h(s) \sse \eta_S(s) \cap (f^d \kcomp h)(s)
\end{equation}
To see that \eqref{eq:congr-star-dual} implies \eqref{eq:congr-dual-cross},
take $h =g^d$.
First, $S \setminus U \notin g(s) \iff U \in g^d(s) = h(s)$.
Second,
$g^d(s) \sse \eta_S(s)$ follows from $\eta_S(s) \sse g(s)$ and items 3 and 6.
Similarly,
$g^d(s) \sse (f^d \kcomp g^d)(s)$ follows from $(f \kcomp g)(s) \sse g(s)$ and items 2, 3 and 6.
The implication  \eqref{eq:congr-dual-cross} $\Rightarrow$ \eqref{eq:congr-star-dual} follows
by a similar argument taking $g=h^d$.
Item 8 is proved in a similar manner as item 7.\qed
\end{proof}
}
\ifWithProofs
\proofLemMFcongruence
\fi

\blue{We now have all the definitions in place to define game models and the semantics of formulas and games. We first give some intuitions. A game model consists of a state space together with interpretations of atomic propositions (as subsets of the state space) and atomic games (as monotone neighbourhood frames). The semantics of complex formulas and complex games is then defined by mutual induction. For a formula $\phi$, the semantics $\sem{\phi}$ is defined via the usual definitions from monotone modal logic. For a game $\alpha$, the semantics $\nsem{\alpha}$ is a monotone neighbourhood frame defined via the game constructions given above. The subsets $U$ in $\nsem{\alpha}(s)$ are the sets of outcomes that Angel can ``force'' when playing the game $\alpha$ in state $s$.}

\begin{definition}\label{def:standard_sem}
A \emph{game model} is a triple $\M = (S,\gamma,\Val)$ where $S$ is a set of states, $\gamma: \AtG \to \MFr(S)$ is a $\AtG$-indexed
collection of monotone neighbourhood frames, which provides an interpretation of atomic games,
and $\Val: \Prop \to \Pow(S)$ is a valuation
of atomic propositions. For $\varphi \in \Form$ and $\alpha \in \GT$ we define the semantics 
$\sem{\varphi}_\M \subseteq S$ and $\nsem{\alpha}_\M \in \MFr(S)$ by induction on the term structure:
\[\begin{array}{l}
\begin{array}[t]{rclclclclclclcl}
\sem{p}_\M & \isdef & \Val(p) \quad \mbox{ for } p \in \Prop &&
\sem{\neg \varphi}_\M & \isdef & S \setminus \sem{\varphi}_\M \\[.2em]
\sem{\varphi_1 \vee \varphi_2}_\M & \isdef & \sem{\varphi_1}_\M \cup \sem{\varphi_2}_\M &&
\sem{\varphi_1 \wedge \varphi_2}_\M & \isdef & \sem{\varphi_1}_\M \cap \sem{\varphi_2}_\M \\[.2em]
\sem{\diam{\alpha} \varphi}_\M & \isdef & \{ s \in S \mid \sem{\varphi}_\M \in \nsem{\alpha}_\M(s) \} && 
\nsem{\alpha;\beta}_\M & \isdef & \nsem{\alpha}_\M \kcomp \nsem{\beta}_\M \\[.2em]
\nsem{g}_\M & \isdef & \gamma(g) \mbox{ for } g \in \AtG &&
\nsem{\alpha^d}_\M & \isdef & (\nsem{\alpha}_\M)^d\\[.2em]
\nsem{\alpha \cup \beta}_\M & \isdef & \nsem{\alpha}_\M \cup \nsem{\beta}_\M &&
\nsem{\alpha \cap \beta}_\M & \isdef & \nsem{\alpha}_\M \cap \nsem{\beta}_\M\\[.2em]
\nsem{\kiter{\alpha}}_\M & \isdef & \kiter{(\nsem{\alpha}_\M)} &&
\nsem{\kcross{\alpha}}_\M & \isdef & \kcross{(\nsem{\alpha}_\M)} \\[.2em]
\nsem{\psi ?}_\M & \isdef & \lambda x. \left\{ \begin{array}{ll}
							\eta_S (x) & \mbox{ if } x \in \sem{\psi}_\M \\
							\emptyset & \mbox{ otherwise.}
							\end{array} \right. &&
\nsem{\bang{\psi}}_\M & \isdef & \lambda x. \left\{ \begin{array}{ll}
							\eta_S (x) & \mbox{ if } x \not\in \sem{\psi}_\M \\
							\Pow S & \mbox{ otherwise.}
							\end{array} \right.
\end{array}
\end{array}\]		
\end{definition}

We write $\phi \equiv \psi$ if for all $\M$, $\sem{\phi}_\M = \sem{\psi}_\M$.
Similarly, we write $\alpha \equiv \beta$ if for all $\M$, $\nsem{\alpha}_\M = \nsem{\beta}_\M$.
We will often omit the subscript $\M$, if $\M$ is clear from the context, or irrelevant.

The following lemma states some basic identities involving the dual operator,
and a congruence property.

\def\stateLemCongruence{
  Let $\phi,\psi \in \Form$ and $\alpha, \beta \in \GT$. We have:\\[.5em]
  \begin{tabular}{lcl}
  1. $(\alpha^d)^d \equiv \alpha$ &&
  2. $(\alpha;\beta)^d \equiv \alpha^d ; \beta^d$ \\ 
  3. $(\alpha \cup \beta)^d \equiv \alpha^d \cap \beta^d$ & \qquad\qquad &
  4. $(\alpha \cap \beta)^d \equiv \alpha^d \cup \beta^d$ \\
  5. $(\iter{\alpha})^d \equiv \diter{(\alpha^d)}$ &&
  6. $(\diter{\alpha})^d \equiv \iter{(\alpha^d)}$ \\
  7. $({\psi}?)^d \equiv \dtest{(\lnot\psi)}$ &&
  8. $(\dtest{\psi})^d \equiv {(\lnot\psi)}?$\\
  9. $\diam{\alpha^d}\phi \equiv \lnot\diam{\alpha}\lnot\phi$ &&
  10. If $\alpha \equiv \beta$ and $\phi \equiv \psi$ then $\diam{\alpha}\phi \equiv \diam{\beta}\psi$ 
  \end{tabular}
}
\begin{lemma}\label{lem:congruence}
\stateLemCongruence
\end{lemma}

\def\proofLemCongruence{
\begin{proof}
  Items 1-6 follow from Lemma~\ref{lem:MF-congruence},
  using the compositional semantics of the game operations.
  We show item 7 (item 8 can be proved similarly).
  \[\begin{array}{lrcl}
\text{Case } s \in \sem{\psi}_\M:\quad &
  U \in \nsem{((\psi)?)^d}(s) & \iff & S\setminus U \notin \nsem{(\psi)?}(s) \\
  && \iff & s \notin S\setminus U \\
  && \iff & s \in U \\
  && \iff & U \in \eta_S(s) = \nsem{\dtest{\lnot\psi}}(s).
\end{array}
\]
\[\begin{array}{lrcl}
\text{Case } s \in \sem{\psi}_\M: \quad &
  U \in \nsem{((\psi)?)^d}(s) & \iff & S\setminus U \notin \nsem{(\psi)?}(s) \\
  && \iff & S\setminus U \notin \emptyset\\
  && \iff & true \\
  && \iff & U \in \Pow(S) = \nsem{\dtest{\lnot\psi}}(s).
\end{array}
\]
Items 9 and 10 are also straightforward to prove using the standard semantics in Definition~\ref{def:standard_sem}.\qed
\end{proof}
}

\ifWithProofs
\proofLemCongruence
\fi

We will make frequent use of the fact that all formulas and game terms can be reduced to a dual and negation normal form.

\begin{definition}
  A formula $\phi \in \Form$, resp.\ game term $\alpha \in \GT$,
  is in \emph{dual and negation normal form (DNNF)}
  if dual is only applied to atomic games and negations occur only in front of proposition letters.
  We denote by $\FormDNNF$ the set of formulas in DNNF,
  and by $\GtDNNF$ the set of game terms in DNNF.
\end{definition}


\def\stateLemDNNF{
  For all $\phi \in \Form$, there is a DNNF formula $\dnnf{\phi}$ such that 
  $\phi \equiv \dnnf{\phi}$.
  For all $\alpha \in \GT$,  there is a DNNF game term $\dnnf{\alpha}$ such that $\alpha \equiv \dnnf{\alpha}$. 
}  
\begin{lemma}\label{lem:DNNF}
\stateLemDNNF
\end{lemma}

\def\proofLemDNNF{
\begin{proof}
  We define $\dnnf{\phi}$ and $\dnnf{\alpha}$ inductively over formulas and game terms.
  On atomic propositions and atomic games, $\dnnf{-}$ acts as identity.
  On formulas with main connective different from $\lnot$ and game terms with main constructor different from $^d$, $\dnnf{-}$ just distributes over the main connective/constructor.
  In particular, $\dnnf{\diam{\alpha}\phi} = \diam{\dnnf{\alpha}}\dnnf{\phi}$.
  For the remaining cases,
  $\dnnf{\phi}$ and $\dnnf{\alpha}$ are defined as follows:
  \[\begin{array}[t]{lclcl}
  \dnnf{\lnot p} &=& \lnot p, & \\
  \dnnf{\lnot\lnot \phi} &=& \dnnf{\phi}, & \\
  \dnnf{\lnot (\phi \land \psi)} &=& \dnnf{\lnot\phi} \lor \dnnf{\lnot\psi}, \\
  \dnnf{\lnot (\phi \lor \psi)} &=& \dnnf{\lnot\phi} \land \dnnf{\lnot\psi}, \\
  \dnnf{\lnot \diam{\alpha}\phi} &=& \diam{\dnnf{\alpha^d}}\dnnf{\lnot\phi},  \\
    \dnnf{({\phi}?)^d} &=& \dtest{\dnnf{\lnot\phi}}, \\
    \dnnf{(\dtest{\phi})^d} &=& {\dnnf{\lnot\phi}}?, \\
  \end{array}\qquad
  \begin{array}[t]{lclc}
    \dnnf{g^d} &=& g^d, \\
    \dnnf{(\alpha^d)^d} &=& \dnnf{\alpha}, \\
    \dnnf{(\alpha \cup \beta)^d} &=& \dnnf{\alpha^d} \cap \dnnf{\beta^d}, \\
    \dnnf{(\alpha \cap \beta)^d} &=& \dnnf{\alpha^d} \cup \dnnf{\beta^d}, \\
    \dnnf{(\alpha ; \beta)^d} &=& \dnnf{\alpha^d} ; \dnnf{\beta^d}, \\
    \dnnf{(\iter{\alpha})^d} &=& \diter{\dnnf{\alpha^d}}, \\
    \dnnf{(\diter{\alpha})^d} &=& \iter{\dnnf{\alpha^d}}, \\
\end{array}
  \]
We prove that $\phi \equiv \dnnf{\phi}$ and $\alpha\equiv \dnnf{\alpha}$
by induction on the length of formulas and game terms, i.e. the number of symbols, not counting parentheses.
  For example the length of $\diam{(p?;g^d)^*}(\lnot p \land q)$ is 10. 

\emph{Base case:} 
For atomic propositions and atomic games (which have length 1), it is trivial.
Similarly for formulas and games of length 2.
So assume the lemma holds for formulas and games of strictly shorter length.

\emph{Step:}
The induction step is straightforward using the IH and Lemma~\ref{lem:congruence}.
For example,
$\dnnf{\lnot\diam{\alpha}\phi} = \diam{\dnnf{\alpha^d}}\dnnf{\lnot\phi} \equiv \diam{\alpha^d}\lnot\phi \equiv \lnot\diam{\alpha}\lnot\lnot\phi \equiv \lnot\diam{\alpha}\phi$ where the second step uses the IH and Lemma~\ref{lem:congruence}(10). Note here that the length of $\alpha^d$ and $\lnot\phi$ are both less than the length of $\lnot\diam{\alpha}\phi$. The third step uses Lemma~\ref{lem:congruence}(9).\qed
\end{proof}
}
\ifWithProofs
\proofLemDNNF
\fi

From now on we will generally assume that formulas are in DNNF.
The following lemma lists some crucial validities that form the basis
for the definition of the game semantics in the next section.
It is straightforward to verify that these formulas are valid. 

\begin{lemma}\label{lem:axioms}
  The following formulas are valid in all game models:\\
  
    $\begin{array}[t]{rclcrcl}
      \diam{\alpha;\beta}\phi & \lra & \diam{\alpha}\diam{\beta}\phi &&
      \diam{\alpha^d}\phi & \lra & \lnot\diam{\alpha}\lnot\phi \\[.2em]
      \diam{\alpha\cup\beta}\phi & \lra & \diam{\alpha}\phi \lor \diam{\beta}\phi & \qquad\qquad&
      \diam{\alpha\cap\beta}\phi & \lra & \diam{\alpha}\phi \land \diam{\beta}\phi \\[.2em]
     \diam{\iter{\alpha}}\phi & \lra &   \phi \lor \diam{\alpha}\diam{\iter{\alpha}}\phi&&
      \diam{\diter{\alpha}}\phi & \lra & \phi \land \diam{\alpha}\diam{\diter{\alpha}}\phi \\[.2em]
      \diam{{\psi}?}\phi & \lra & \psi \land \phi &&
      \diam{\dtest{\psi}}\phi & \lra & \psi \lor \phi 
    \end{array}$
      
\end{lemma}


\section{Game Semantics for Game Logic}\label{sec:games}

In this section we will see how games provide an operational semantics
for game logic. In particular, we will develop a two-player evaluation
game for game logic, very much in the spirit of
Berwanger~\cite{Berwanger:GL-parity}. Note however, that the ambient
model-theoretic structures in our setting are \emph{monotone
neighbourhood structures}, whereas Berwanger restricts to (relational)
Kripke structures. Our approach allows for a neat formulation of some
useful additional observations involving the unfolding games related to
monotone operations on full powersets~\cite{vene12:lect}.

\subsection{Game Preliminaries}

Two-player \emph{graph games} are an important tool for fixpoint logics.
We will briefly recall their definition and the related terminology.
For a more comprehensive account of these games, the reader is referred 
to~\cite{grae:auto02}.
%
A graph game is played on a \emph{board} $B$, that is, a set of \emph{positions}.
Each position $b \in B$ \emph{belongs} to one of the two \emph{players}, \Heloise (abbr.~$\E$) and 
\Abelard (abbr.~$\abel$). Formally we write $B = B_{\E} \cup B_{\abel}$, and for each position $b$ we 
use $P(b)$ to denote the player $i$ such that $b \in B_{i}$.
Furthermore, the board is endowed with a binary relation $E$, so that each
position $b \in B$ comes with a set $E[b] \sse B$ of \emph{successors}. Note
that we do not require the games to be strictly alternating, i.e., successors of
positions in $B_\E$ or $B_\abel$ can lie again in $B_\E$ or $B_\abel$, respectively.
Formally, we say that the \emph{arena} of the game consists of a directed
two-sorted graph $\bbB = (B_{\E} ,  B_{\abel}, E)$.

A \emph{match} or \emph{play} of the game consists of the two players moving
a pebble around the board, starting from some \emph{initial position} $b_{0}$.
When the pebble arrives at a position $b \in B$, it is player $P(b)$'s turn
to move; (s)he can move the pebble to a new position of their liking, but 
the choice is restricted to a successor of $b$.
Should $E[b]$ be empty then we say that player $P(b)$ \emph{got stuck} at
the position.
A \emph{match} or \emph{play} of the game thus constitutes a (finite or 
infinite) sequence of positions $b_{0}b_{1}b_{2}\ldots\ $ such that 
$b_{i}Eb_{i+1}$ (for each $i$ such that $b_{i}$ and $b_{i+1}$ are defined).
A \emph{full play} is either (i) an infinite play or (ii) a finite play in
which the last player got stuck.
A non-full play is called a \emph{partial} play. 
Each full play of the game has a \emph{winner} and a \emph{loser}.
A finite full play is lost by the player who got stuck; the winning condition
for infinite games is usually specified using 
a so-called \emph{parity function}, ie., a function $\Om: B \to \mathbb{N}$ that maps each 
position to a natural number (its \emph{priority}) and that has
finite range. An infinite play $\play = b_0 b_1 \dots b_n \dots \in B^\omega$
is won by \Heloise if $\max \{ \Om(b) \mid b \in \mathrm{Inf}(\play) \}$ is even, where
$\mathrm{Inf}(\play)$ denotes the positions from $B$ that occur infinitely often in $\play$. Otherwise
\Abelard wins this play. A graph game with parity function $\Omega$ is a \emph{parity game}.
All graph games used in this paper are parity games, but we will not specify the parity function
explicitly in simple cases (e.g. when one of the players is supposed to win all infinite plays).
%


A \emph{strategy} for player $i$ tells player $i$ how to play at all positions where it is $i$'s turn to move. 
A strategy can be represented as a \emph{partial} function which maps partial
plays 
$\beta = b_{0}\cdots b_{n}$ with $P(b_{n}) = i$ to legal next positions
(that
is, to elements of $E[b_{n}]$), and which is undefined for partial plays
$\beta = b_{0}\cdots b_{n}$ with $E[b_{n}] = \emptyset$.
We say that a play $\play = b_1 \dots b_n \dots \in B^* \cup B^\omega$
follows a strategy $f$ if for all positions $b_j$ in $\play$ on which $f$ is defined
we have $f(b_j) = b_{j+1}$.
A strategy is \emph{positional} if it only depends on the current 
position of the match. 
A strategy is \emph{winning for player $i$} from position $b \in B$ if it
guarantees $i$ to win any match with initial position $b$, no matter how
the adversary plays --- note that this definition also applies to positions
$b$ for which $P(b) \neq i$.
A position $b \in B$ is called a \emph{winning position} for player $i$, if 
$i$ has a winning strategy from position $b$; the set of winning positions
for $i$ in a game $\game$ is denoted as $\Win_{i}(\game)$.
Parity games are \emph{positionally determined}, i.e., at each position
of the game board exactly one of the players has a positional winning strategy
(cf.~\cite{most:game91,emer:tree91}).

\subsection{Definition of the Evaluation Game}
In order to be able to trace the unfoldings of fixpoint operators within games we need some terminology concerning the nesting of fixpoints. Firstly, we need notation for the subterm relation and the definition of a parity map for a formula.

\begin{definition}
	We let $\subf \subseteq (\Form \cup \GT)^2$ be the subterm relation
	on formulas and game terms, ie., $\xi_1 \subf \xi_2$ if either $\xi_1 = \xi_2$
	or $\xi_1$ is a proper subterm of $\xi_2$.
\end{definition}
	


\begin{definition} \label{omega on fixpoints}
	For a term $\xi \in \Form \cup \GT$ we let 
	$\Fix(\xi) \isdef \{ \iter{\alpha} \mid \alpha\in \GT, \iter{\alpha} \subf \xi \} \cup 
	\{ \diter{\alpha} \mid \alpha\in \GT, \diter{\alpha} \subf \xi\}$.
	A parity function for a formula $\varphi$ in DNNF is a partial map $\Om: \Fix(\varphi) \to \omega$ such that
	\begin{enumerate}
		\item $\alpha_1 \subf \alpha_2$ implies $\Om(\alpha_1) < 
			\Om(\alpha_2)$ for all $\alpha_1, \alpha_2 \in \Fix(\varphi)$ with $\alpha_1 \not= \alpha_2$, and
		\item for all $\alpha\in \Fix(\varphi)$, $\Om(\alpha)$ is even iff $\alpha= \diter{\rho}$ is a demonic iteration.
	\end{enumerate}
	We define the canonical parity function $\Om_\can: \Fix(\varphi) \to \omega$ associated with $\varphi$ as the partial function given by $\Om_\can(\iter{\alpha}) = 2n+1$ and $\Om_\can(\diter{\alpha}) = 2n$ where
	$n = \card{\Fix(\iter{\alpha})}$ and $n = \card{\Fix(\kcross{\alpha})}$, respectively.
	\blue{The \red{canonical} parity function formalises the fact that any fixpoint operator dominates any other fixpoint operator in its scope.} 
\end{definition}

\begin{definition}\label{def:game_sem}
Let $\M=(S,\gamma,\Val)$ be a  game model, let $\varphi \in \Form$ be a formula in DNNF and let $\Om: \Fix(\varphi) \to \omega$ be a parity function for $\varphi$. We define the evaluation game $\Ev(\M,\varphi)$ as the parity graph game with the game board specified in Fig.~\ref{fig:evgame} and the parity function $\Om_\Ev$ given by 
\[ \Om_\Ev(b) \isdef \left\{ \begin{array}{ll}
	 \Om(\alpha) & \mbox{ if } b=(x,\diam{\alpha}\psi) \mbox{ for some } \alpha\in \Fix(\varphi)  \\
	 0 & \mbox{ otherwise.} 
	 	 \end{array} \right.
\]
\end{definition}
\nopagebreak
\begin{figure}[!h]
\begin{center}
\begin{tabular}{ccc}
\textbf{Formula Part} & \quad & \textbf{Game Part}
 \\
\begin{tabular}[t]{|l|c|l|}
\hline
Position $b$ & P(b) & Moves E[b] \\
\hline\hline
$(s,p), s \in \Val(p)$ & \Ab & $\emptyset$ \\
\hline
$(s,p), s \notin \Val(p)$ & \He & $\emptyset$ \\
\hline
$(s,\neg p), s \in \Val(p)$ & \He & $\emptyset$ \\
\hline
$(s,\neg p), s \not\in \Val(p)$ & \Ab & $\emptyset$ \\
\hline

$(s,\varphi \land \psi)$ & \Ab & $\{(s,\varphi),(s,\psi)\}$ \\
\hline

$(s,\varphi \lor \psi)$ & \He & $\{(s,\varphi),(s,\psi)\}$ \\
\hline

$(s,\diam{g} \varphi)$ & \He & $\{(U,\diam{g}\varphi) \mid U \in
\nsem{g}(s) \}$ \\
\hline
$(U,\diam{g} \varphi)$ & \Ab & $\{(s,\varphi) \mid s \in U\}$ \\
\hline
$(s,\diam{g^d} \varphi)$ & \Ab & $\{(U,\diam{g^d}\varphi) \mid U \in
\nsem{g}(s) \}$ \\
\hline
$(U,\diam{g^d} \varphi)$ & \He & $\{(s,\varphi) \mid s \in U\}$ \\
\hline
\end{tabular} & &
\begin{tabular}[t]{|l|c|l|}
\hline
Position $b$ & P(b) & Moves E[b] \\
\hline\hline
$(s,\diam{\alpha\comp\beta}\varphi)$ & $\star$ &
$\{(s,\diam{\alpha}\diam{\beta}\varphi)\}$ \\
\hline
$(s,\diam{\alpha\cup\beta}\varphi)$ & $\star$ &
$\{(s,\diam{\alpha}\varphi \lor \diam{\beta}\varphi)\}$ \\
\hline
$(s,\diam{\alpha\cap\beta}\varphi)$ & $\star$ &
$\{(s,\diam{\alpha}\varphi \land \diam{\beta}\varphi)\}$ \\
\hline

$(s,\diam{\alpha^*}\varphi)$ & $\star$ &
$\{(s,\varphi \lor \diam{\alpha}\diam{\alpha^*}\varphi)\}$ \\
\hline

$(s,\diam{\alpha^\cross}\varphi)$ & $\star$ &
$\{(s,\varphi \land \diam{\alpha}\diam{\alpha^\cross}\varphi)\}$ \\
\hline

$(s,\diam{\psi?}\varphi)$ & $\star$ &
$\{(s,\psi \land \varphi)\}$ \\
\hline

$(s,\diam{\bang{\psi}}\varphi)$ & $\star$ &
$\{(s,\psi \lor \varphi)\}$ \\
\hline
\end{tabular} 
\end{tabular}
\end{center}
\caption{Game board of the evaluation game. We use $P(b) = \star$ to express that it is irrelevant which player moves, since there is exactly one possible move.}\label{fig:evgame}
\end{figure}



\subsection{Adequacy of Game Semantics}
In this section we show that the game semantics of Definition~\ref{def:game_sem} is equivalent to the standard 
semantics of game logic from Definition~\ref{def:standard_sem} where we assume w.l.o.g. that formulas are in DNNF.

To compare the two different semantics we need a game characterisation of the $\kiter{(\_)}$ and $\kcross{(\_)}$-operations. As both operations are defined as fixpoints they can be characterised via fixpoint games \blue{(these games are straightforward adaptation of the unfolding game described in~\cite{vene12:lect})}.
We provide some intuition below the definition.

\begin{definition} \label{fixpoint game}
    Let $\alpha\in \GT$ be a game term, let $\M=(S,\gamma,\Val)$ be a  game model and let 
    $U \subseteq S$. The games
    $\game(\M,{\kiter{\alpha}},U)$ and $\game(\M,{\kcross{\alpha}},U)$ have the following game boards:
    \begin{center}
      \begin{tabular}{cc}
        Board of $\game(\M,{\kiter{\alpha}},U)$: &
        Board of $\game(\M,{\kcross{\alpha}},U)$: \\
    \begin{tabular}{|c|c|c|} 
    	\hline
    	Pos. $b$ &  $P(b)$ & Moves $E[b]$  \\
    	\hline \hline 
		$s \in S$ & \He & $ \left\{ \begin{array}{ll} 
					\{ \emptyset \}  & \red{\mbox{ if } s \in U} \\
					\nsem{\alpha}(s) & \mbox{ otherwise.}  \\
					\end{array} \right.$ \\
		\hline
		$U' \in \Pow(S)$ & \Ab & $U'$ \\
		\hline
\end{tabular}         
&
    \begin{tabular}{|c|c|c|}
    	\hline
    	    	Pos. $b$ & $P(b)$ & Moves $E[b]$  \\
 		\hline \hline 
    	 $s \in S$ & \He & $ \left\{ \begin{array}{ll} 
					\nsem{\alpha}(s) & \mbox{ if } s \in U \\
					\emptyset & \mbox{ otherwise.}  \\
					\end{array} \right.$ \\
		\hline
		$U' \in \Pow(S)$ & \Ab & $U'$ \\
		\hline
    \end{tabular}
\end{tabular}
      
 \end{center}
    The winning conditions in these games are as usual: finite complete plays are lost by the player that 
    gets stuck. Infinite plays of $\game(\M,{\kiter{\alpha}},U)$ and $\game(\M,{\kcross{\alpha}},U)$ 
    are won by \Abelard and \Heloise, respectively.  
\end{definition}

\red{The fixpoint game $\game(\M,{\kiter{\alpha}},U)$ works as follows.
The objective of \Heloise is to reach $U$ in finitely many rounds of $\alpha$.
At a position $s \in U$, \Heloise can win by choosing the move $\emptyset$ which causes
\Abelard to get stuck in the next step, since he must choose from the empty set of moves.
At a position $s \not\in U$, \Heloise chooses an $\alpha$-neighbourhood $U'$ of $s$,
and in the next step \Abelard then chooses a state $s' \in U'$, and the game continues.
In the game $\game(\M,{\kcross{\alpha}},U)$, the objective of \Heloise is to stay in $U$ indefinitely.
At a position $s \notin U$, she therefore loses immediately (indeed, she is stuck at such positions, since her set of moves is empty).
But at a position $s \in U$, the players play another round of $\alpha$, and the game continues.}

\begin{lemma}\label{lem:fixgames}
  For all $\M = (S,\gamma,\Val)$, $\alpha\in \GT$, $s \in S$ and $U \subseteq S$, we have:\\
  
  \begin{tabular}{lcl}
	$s \in \Win_\E(\game(\M,{\kiter{\alpha}},U))$ & \mbox{ iff } & $U \in \nsem{\kiter{\alpha}}(s)$, and\\
    $s \in \Win_\E(\game(\M,{\kcross{\alpha}},U))$ & \mbox{ iff } & $U \in \nsem{\kcross{\alpha}}(s)$.
  \end{tabular}
\end{lemma}
	The lemma easily follows because 
	the games $\game(\M,{\kiter{\alpha}},U)$ and $\game(\M,{\kcross{\alpha}},U)$
	are instances of Tarski's fixpoint games that characterise least and greatest fixpoints
	of a monotone operator. 

%
The following technical lemma demonstrates that winning strategies for \Heloise 
in the evaluation game entail the existence of certain neighbourhood sets in the game model that 
witness the truth of a modal formula. \blue{There is no requirement on the witness to be non-empty,
  e.g., $s \models \diam{\alpha}\bot$ if $\emptyset \in \nsem{\alpha}(s)$.}
 
\def\stateLemFromEtoG{
	Let $\varphi \in \Form$, let $\M=(S,\gamma,\Val)$ be a game model and consider
	the game $\Ev = \Ev(\M,\varphi)$.
        Assume that $f_\E$ is a winning strategy for \Heloise in $\Ev$, and
        that $(s,\diam{\alpha}\psi) \in Win_\E(\Ev)$.
        Let $\Win_\psi(\Ev) \isdef \{s' \in S \mid 
	(s',\psi) \in \Win_\E(\Ev)\}$ and suppose $\Win_\psi(\Ev) \subseteq \sem{\psi}$. Then 
	(i) $\Suc_{f_\E}(s,\diam{\alpha}\psi) \in \nsem{\alpha}(s)$, and
	(ii) $\Win_\psi(\Ev) \in \nsem{\alpha}(s)$.
}
\def\stateLemFromEtoGbody{
	Let $\varphi \in \Form$, let $\M=(S,\gamma,\Val)$ be a game model and consider
	the game $\Ev = \Ev(\M,\varphi)$.
        Assume that $f_\E$ is a winning strategy for \Heloise in $\Ev$, and
        that $(s,\diam{\alpha}\psi) \in Win_\E(\Ev)$.
        Let $\Win_\psi(\Ev) \isdef \{s' \in S \mid 
	(s',\psi) \in \Win_\E(\Ev)\}$ and suppose $\Win_\psi(\Ev) \subseteq \sem{\psi}$. Then 
	$\Win_\psi(\Ev) \in \nsem{\alpha}(s)$.
}
\begin{lemma}\label{lem:fromEtoG}
\stateLemFromEtoGbody
\end{lemma}

\def\proofLemFromEtoG{
\begin{proof}
	The second item follows easily from the first, as all states that are reachable via some
	winning strategy of \Heloise from a winning position must again be winning positions.
        In other words, $\Suc_{f_\E}(s,\diam{\alpha}\psi) \subseteq \Win_\psi(\Ev)$, hence by monotonicity (i) implies (ii).
	The first item can be proven by
	induction on the structure of $\alpha$. We only discuss the fixpoint operators and angelic tests. All other cases are a matter of routine checking.
	\begin{description}
		\item[Case] $\alpha= \kiter{\alpha'}$. By our assumption \Heloise has a winning strategy at position
		$(s,\diam{\alpha}\psi)$ in $\Ev$. We show that she also has a winning strategy in the game
		$\game = \game(\M,{\kiter{\alpha}},\Suc_{f_\E}(s,\diam{\alpha}\psi))\in \nsem{\alpha}(s)$ at position $s$.
		To see this we equip \Heloise with a strategy in $\game$ such that for any play 
		\[ s U_1 s_1 \dots U_n s_n \]
		there is a ``shadow play'' in $\Ev$ of the form 
		\[ (s,\diam{\kiter{\alpha'}}\psi) \dots (s_1,\diam{\kiter{\alpha'}}\psi) \dots (s_n,\diam{\kiter{\alpha'}}\psi) \]
		that follows \Heloise's winning strategy in $\Ev$. Suppose this connection has been 
		established for $n$ rounds (ie. for all plays of $\game$ of where \Heloise moved at most $n$ times) we are going to describe how to extend it to games with $n+1$ rounds: Consider a $\game$-play of $n$ rounds ending in position 
		$s_n$ and assume that \Heloise has played according to the strategy that we are providing for her. 
		By assumption there is an $\Ev$-play according to \Heloise's winning strategy in $\Ev$ that
		reaches state $(s_n,\diam{\kiter{\alpha'}}\psi)$. 
		\begin{description}
			\item[Subcase] \Heloise's $\Ev$-strategy $f_\E$ requires a move to $(s_n, \psi)$. In this case 
			we have $s_n \in \Suc_{f_\E}(s,\diam{\kiter{\alpha'}}\psi) \subseteq \Win_\E$. 
			Therefore the $\game$-play ending
			in $s_n$ is complete and winning for \Heloise as required.
			\item[Subcase] \Heloise's $\Ev$-strategy requires to move to $(s_n,\diam{\alpha'}\diam{\kiter{\alpha'}}\psi)$. 	As \Heloise's strategy is winning we have $(s_n,\diam{\alpha'}\diam{\kiter{\alpha'}}\psi) \in \Win_\E$ and
			by I.H. we get $U_{n+1} \isdef \Suc_{f_\E}((s_n,\diam{\alpha'}
			\diam{\kiter{\alpha'}}\psi) \in \nsem{\alpha'}(s_n)$.
			We define \Heloise's strategy in $\game$ to move from 
			$s_n$ to $U_{n+1}$. If $U_{n+1} = \emptyset$ we have
			that the $\game$-play is complete and \Heloise wins as required. Otherwise \Abelard picks some 
			$s_{n+1} \in U_{n+1}$. By the definition of $U_{n+1}$ this implies that the $\Ev$-shadow play
			can be prolonged from $(s_n, \diam{\kiter{\alpha'}}\psi)$ to $(s_{n+1},\diam{\kiter{\alpha'}}\psi)$	
			following \Heloise's winning strategy $f_\E$.	
		\end{description} 
		In either subcase we showed how \Heloise can prolong the $\game$-play - unless \Abelard gets stuck - 
		such that there is a suitable
		parallel $\Ev$-play that follows \Heloise's winning strategy $f_\E$. Furthermore, by Def.~\ref{def:suc}
		of $\Suc$ we have that $\kiter{\alpha}$ is the fixpoint with the highest priority 
		that is unfolded within this parallel
		 $\Ev$-play. This means - as the $\Ev$-play is winning for \Heloise - the parallel $\Ev$-play has to stop eventually which implies that \Abelard has to get eventually stuck in the $\game$-play as well. In other words, \Heloise wins each $\game$-play that starts in $s$ and that is played according to the strategy that we devised for her.
		 Thus $U \in \nsem{\kiter{\alpha'}} (s)$ as required.
		\item[Case] $\alpha= \kcross{\alpha'}$. The proof for this case is analogous to the previous one.
		\item[Case] $\alpha= \xi?$. By assumption \Heloise has a winning strategy at $\diam{\xi?}\psi$
		    and we have $\Win_\psi \subseteq \sem{\psi}$.
		    As $(s,\diam{\xi?}\psi)$ is winning for \Heloise, we also have
		    $(s,\psi) \in \Win_\E$ and thus $s \in \Win_\psi \subseteq \sem{\psi}$.
		    Therefore $\nsem{\xi?}(s) = \eta_S(s)$ and hence $\Suc_{f_\E}(s,\diam{\xi?}\psi) = \{s\} \in \nsem{\xi?}(s)$ as required. \qed
	\end{description} 
\end{proof}
}
\ifWithProofs
\proofLemFromEtoG
\fi

The lemma is the key to prove one direction of the adequacy of our game semantics.

\begin{proposition}\label{prop:adequate_first}
	Let $\varphi \in \Form$, let $\M=(S,\gamma,\Val)$ be a game model and consider
    $\Ev = \Ev(\M,\varphi)$. For all $\psi$ occurring in $\Ev$
	we have $\Win_\psi(\Ev) \subseteq \sem{\psi}_\M$.
\end{proposition}
	The claim is proven by induction on $\psi$ and follows easily from
	Lemma~\ref{lem:fromEtoG}.
For the second half of the adequacy theorem we again need a technical lemma.

\def\stateLemReachPsi{
	Let $\M=(S,\gamma,\Val)$ be a game model and let $\varphi \in \Form$. For any position
	$(s,\diam{\alpha}\psi)$ of the game $\Ev=\Ev(\M,\varphi)$ 
	and for all $U \subseteq \sem{\psi}_\M$ with $U \in \nsem{\alpha}(s)$ 
	\Heloise has a strategy
	$f_\E$ such that
	for each finite $\Ev$-play $\play$ starting at $(s,\diam{\alpha}\psi)$ and following $f_\E$ 
	either \Abelard gets stuck or $\play$ reaches a state $(s',\xi') \in S \times \Form$ 
	that satisfies one
	of the following conditions:
	(i) $\xi' \subf \alpha$ and $s' \in \sem{\xi'}$, or  
	(ii) $\xi' = \psi$ and $s' \in U$.
}

\begin{lemma}\label{lem:reachpsi}
\stateLemReachPsi
\end{lemma}

\def\proofLemReachPsi{
\begin{proof}
	The claim is proven by induction on $\alpha$.
	\begin{description}
		\item[$\alpha= g$] If 
		   $U \in \nsem{g}(s)$ for some $U \subseteq \sem{\psi}$, \Heloise's strategy consists of moving 
		   to $U$, which obviously fulfils the conditions of the lemma.
		\item[$\alpha=\alpha_1;\alpha_2$] Let 
		 $U \in\nsem{\alpha_1;\alpha_2}(s)$ with $U \subseteq \sem{\psi}$. This implies $U \in		
		 (\nsem{\alpha_1} \kcomp \nsem{\alpha_2})(s)$, ie., $U' \isdef 
		 \{ s' \in S \mid U \in \nsem{\alpha_2}(s')\} \in \nsem{\alpha_1}(s)$.
		  	It is easy to see that $U' \subseteq \sem{\diam{\alpha_2}\psi}$.
			\Heloise has to move in $\Ev$ 
			from position $(s,\diam{\alpha_1;\alpha_2})$ to $(s,\diam{\alpha_1}\diam{\alpha_2}\psi)$. To the latter position
			we can apply the I.H. which yields a strategy $g$ for \Heloise such that:
		    \begin{itemize}
		    	\item Any play following $g$ starting in $(s,\diam{\alpha_1}\diam{\alpha_2}\psi)$ either leads to some 
					position $(s',\diam{\alpha_2}\psi)$ with $s' \in U'$ 
					or
					a position $(s',\xi)$ with $s' \in \sem{\xi}$ and $\xi \subf \alpha_1$ or
					\Abelard gets stuck before any of the above happens.
				\item Applying for each state of the form $\sigma = (s',\diam{\alpha_2}\psi)$ with 
				    $s' \in U'$ the I.H. yields 
					a strategy $g_\sigma$ such that each $\Ev$-play that starts at $(s',\diam{\alpha_2}\psi)$
					and that follows $g_\sigma$ ends in a state of the form $(s'',\psi)$ with
					$s'' \in \sem{\psi}$ or to a state of the form $(s'',\xi')$ with $s'' \in \sem{\xi'}$
					or Abelard gets stuck.
		    \end{itemize}
		    	These facts show that a combination of the $g_\sigma$ strategies with $g$ 
				 will describe a suitable strategy $f_\E$
				for \Heloise at position $(s,\diam{\alpha_1;\alpha_2}\psi)$.
		\item[$\alpha= g^d$] If $U \in \nsem{g^d}(s)$ for some $U \subseteq \sem{\psi}$ then for
			all $U' \in \nsem{g^d}(s)$ we have $U' \cap U \not=\emptyset$.
			For any possible choice $U' \in \nsem{g^d}(s)$ by \Abelard we define \Heloise's response
			according to her strategy $f_\E$ to be some element
			of $U \cap U'$.
		\item[$\alpha= \xi?$] If $U \in \nsem{\xi?}(s)$ for some $U \subseteq \sem{\psi}$  we have
		     $s \in U$  
		     and \Heloise will move to position $(s,\psi)$. 
		\item[$\alpha= \bang{\xi}$] If $U \in \nsem{\bang\xi}(s)$ for some $U \subseteq \sem{\psi}$
		     then either 
			 $s \in \sem{\xi}$ of $s \in U$.
			 In the first case \Heloise's strategy is to move to
			 $(s,\xi)$ 
			 In case $s \in U$, \Heloise's strategy is to move to
			 $(s,\psi)$. 
		
		\item[$\alpha= \kiter{\alpha'}$] Consider some $U \in \nsem{\kiter{\alpha'}}(s)$ with 
			$U \subseteq \sem{\psi}$.
			Thus \Heloise has a winning strategy in $\game = \game(\M,\kiter{\alpha'},U)$ at position $s$.
			Suppose her strategy in $\game$ is to move to some $U_\exists \subseteq S$. 
			\begin{description}
				\item[Subcase] If $s \in U$ we can assume that w.l.o.g. $U_\E = \emptyset$. Then by
				I.H. there is a strategy for \Heloise in $\Ev$ such that in any play following that strategy
				either \Abelard gets stuck or a position $(s',\xi')$ is reached with $\xi' \subf \alpha'$
				and $s' \in \sem{\xi'}$. As $\xi' \subf \diam{\kiter{\alpha'}}\psi$ this case meets the requirements of the lemma.
				\item[Subcase] Otherwise we know that
				$U_\E \in \nsem{\alpha'}(s)$ and - as \Heloise's strategy in $\game$ is winning -
				we can assume that $U_\E \subseteq \Win_\E(\game) = U$.
				Furthermore
				$U \in \nsem{\alpha'} \kcomp \nsem{\kiter{\alpha'}}(s)$ 
				and hence $V \isdef \{ s' \in S \mid U \in \nsem{\kiter{\alpha'}}(s') \} \in \nsem{\alpha'}(s)$.
				Because $V \subseteq \{ s' \in S \mid \sem{\psi} \in \nsem{\kiter{\alpha'}}(s') \}$ 
				we have $V \subseteq \sem{\diam{\kiter{\alpha'}}\psi}$.
				This implies that $U \subseteq V$ as $s' \in U$ implies
				$U \in \nsem{\kiter{\alpha'}}(s')$.
			
				Collectively this shows that $U_\E \in \nsem{\alpha'}(s)$ and
				 $U_\E \subseteq U \subseteq V \subseteq \sem{\diam{\kiter{\alpha'}}\psi}$, ie, we can apply the I.H. 
				 to $U_E$ and $\diam{\kiter{\alpha'}}\psi$.
				
				We obtain a strategy $g$ for \Heloise in $\Ev$ such that in any play $\play$
				following $g$
				starting at $(s,\diam{\alpha'}\diam{\kiter{\alpha'}}\psi)$ 
				either \Abelard gets stuck or 
				$\play$ reaches a position $(s',\xi')$ with $s \in \sem{\xi'}$  (in these two
				cases we do not continue the play as the conditions of the lemma are met) or 
				such that $\play$ reaches a position $(s'',\diam{\kiter{\alpha'}}\psi)$ with $s'' \in U_\E$.
				In the latter case we note that there is a shadow play $s \to U_\E \to s''$ of 
				$\game$ that follows \Heloise's winning strategy and thus we can apply
				the same reasoning that we applied to $(s,\diam{\kiter{\alpha'}}\psi)$ to the new state $(s'',\diam{\kiter{\alpha'}}\psi)$ and so forth. 			\end{description}
				This implies that the second subcase can only
				occur finitely often as any corresponding $\game$-play has to be won be \Heloise and thus needs
				to be finite. In other word, after finitely many iterations of our argument, either \Abelard will get stuck or the play will reach a position of the form $(s''',\psi)$ with $s''' \in \sem{\psi}$ as required.
		\qed
	\end{description} 
\end{proof}
}
\ifWithProofs
\proofLemReachPsi
\fi

\def\statePropEvalGame{
	Let $\M = (S,\gamma,\Val)$ be a  game model and consider the game $\Ev = \Ev(\M,\varphi)$ for some
	$\varphi \in \Form$. There is a strategy $f_\E$ for \Heloise that is winning for \Heloise
	for all game positions $(s,\psi)$ such that $s \in \sem{\psi}_\M$.
}  
\begin{proposition}\label{prop:eval-game}
\statePropEvalGame
\end{proposition}

\def\proofPropEvalGame{
\begin{proof}
	The claim is proven by induction on the order $\subf$ on formulas. We provide the details for some important
	cases:
	\begin{description}
		\item[$\psi = p$] In this case $s \in \sem{p}$ implies $s \in \Val(p)$ and
			\Heloise wins the play immediately. 
		\item[$\psi = \neg p$] similar to previous case.
		\item[$\psi = \psi_1 \land \psi_2$] In this case $s \in \sem{\psi_1 \land \psi_2}$ means that
		  $s \in \sem{\psi_i}$ for $i =1,2$. A move of \Abelard from $(s,\psi)$ to $(s,\psi_i)$ will be answered 
		  by \Heloise following her winning strategy $f_\E$ that is defined at $(s,\psi_i)$ by the I.H. 
		  \\ \vdots
		\item[$\psi = \diam{\alpha} \xi$] In this case, by Lemma~\ref{lem:reachpsi}, \Heloise has a strategy $f$
		   such that each play following $f$ eventually reaches a position of the form $(s',\xi')$ with
		   $s' \in \sem{\xi'}$, $\xi' \not= \psi$ and $\xi' \subf \psi$ or \Abelard gets stuck. We define \Heloise's strategy at $(s,\diam{\alpha}\xi)$ to be $f$ until in the resulting play one of those two possibilities happens: If \Abelard gets stuck, \Heloise wins the play and we are done. If the play reaches a position of the form $(s',\xi')$ with $s' \in \sem{\xi'}$ we let \Heloise
		   continue to play her winning strategy $f_\E$ that exists by I.H.  \qed

	\end{description}
	\end{proof}
}
In summary, Proposition~\ref{prop:adequate_first} and Proposition~\ref{prop:eval-game} imply that our game semantics for game logic is adequate:
\begin{theorem}\label{thm:eval-game}
	Let $\M = (S,\gamma,\Val)$ be a  game model and consider the game $\Ev = \Ev(\M,\varphi)$ for some
	$\varphi \in \Form$. Then for all positions $(s,\psi)$ in $\Ev$ we have $(s,\psi) \in \Win_\E(\Ev)$ iff
	$\M,s \models \psi$.
\end{theorem}
\ifWithProofs
\proofPropEvalGame
\fi

\section{Syntax Graphs}
\label{sec:graphs}

In this section we introduce syntax graphs which we then use later to
provide an automata-theoretic characterisation of game logic. Syntax
graphs are a generalisation of syntax trees that allow cycles and
sharing of subterms. Another perspective is that they are a graph-based
description of the alternating tree automata from
\cite{Wilke01:ATA,Kirsten02:ATA}. We discuss the precise connection
after the definition of syntax graphs and their game semantics.

\subsection{Graph Basics}
\label{graphs}

We first recall some basic notions and fix notation.
A graph is a pair $\G = (V,E)$ where $V$ is a set of vertices $V$ and
$E \subseteq V \times V$ is a set of edges. We will use the following notation:
$vEw$ iff $(v,w) \in E$ iff $w \in E(v)$, and call $w$ a successor of $v$.

Let $\G=(V,E)$ be a graph.
A \emph{path} $p$ in $\G$ is a sequence of
vertices $p = v_1 \dots v_n$ such that $v_i E v_{i+1}$ for all $i < n$.
We say that $v_n$ is \emph{reachable} from $v_1$ if a path $p = v_1 \dots v_n$ exists.
Note that every vertex is always reachable from itself.
A \emph{cycle} $c = v_1 \dots v_n$ is any path such that
$v_1 = v_n$ and $n \geq 2$.

A path $p = v_1 \dots v_n$ is \emph{simple} if all the $v_i$ for $i \leq n$
 are distinct. A cycle $c = v_1 \dots v_n$ is
\emph{simple} if all the $v_i$ for $i < n$ are distinct
Every path can be \emph{contracted} to a simple path with the same start
and end points, To see how this works consider a path $p$ that contains
a repetition of some vertex $u \in V$. This means that $p$ is of the
form $p = q u m u r$, for paths $q$, $m$ and $r$. We contract $p$ to the
path $q u r$ with the same starting and end points, in which there is
one less occurrence of $u$. We can repeat this procedure until we obtain
a simple path.

A \emph{pointed} graph $\G = (V,E,v_I)$ is a graph $(V,E)$ together
with a $v_I \in V$ that we call the \emph{initial} vertex of $\G$.
If $\G$ is a graph $(V,E)$ or a pointed graph $(V,E,v)$
and $v_I$ is a vertex in $\G$, we
define $\reinit{\G}{v_I} = (V',E',v_I)$ to be the
\emph{subgraph generated by $v_I$ in $\G$},
i.e., $V'$ is the set of vertices that are reachable from $v_I$
and $E' = E \cap (V' \times V')$.

A pointed graph $\G = (V,E,v_I)$ is \emph{reachable} if every $v \in V$ is
reachable from $v_I$. Note that $\reinit{\G}{v_I}$ is
always reachable.

\subsection{Syntax Graphs}
\label{syntax graphs}

We define the following sets of label symbols:
$\Lit = \Lbl{0} \isdef \{ p, \lnot p \mid p \in \Prop\}$,
$\Latt = \Lbl{2} \isdef \{\land,\lor\}$ and
$\Mod = \Lbl{1} \isdef \{\diam{g} \mid g \in \AtomicGames\} \cup \{\diam{g^d} \mid g \in \AtomicGames\}$.
The labels $\Lbl{0}, \Lbl{1}, \Lbl{2}$ can be given an arity in the expected manner,
namely, for $l \in \Lbl{i}$, $\arity(l) = i$.
We let $\Lbl{} \isdef \Lbl{0} \cup\Lbl{1} \cup \Lbl{2}$.

\begin{definition}
A \emph{syntax graph} $\G = (V,E,L,\Omega)$ is a finite graph $(V,E)$
together with a \emph{labelling function}
$L : V \to \Lbl{}$ and a partial \emph{priority
function} $\Omega : V \pfun \omega$ satisfying the following two conditions:
\begin{description}
\item[(arity condition)]
  For all $v \in V$, 
  $|E(v)| = \arity(L(v))$.
 \item[(priority condition)] On every simple cycle of $(V,E)$ there is at least one vertex on
which $\Omega$ is defined.
\end{description}
\end{definition}


Later we will show that formulas correspond to syntax graphs, and
game terms correspond to syntax graphs with a special atomic proposition
that marks an ``exit'' from the graph. The idea is that a game term $\alpha$
is viewed as the modality $\diam{\alpha}$ which still needs a formula $\phi$ in order to become a
formula $\diam{\alpha}\phi$, and an exit marks a place in the graph where $\phi$ can be inserted.

\begin{definition} \label{exit}
  A proposition letter $e$ is an \emph{exit} of a syntax graph
  $\G = (V,E,L,\Omega)$ if there is a vertex $v \in V$ with $L(v)=e$
  and there is no $v \in V$ with $L(v) = \neg e$.

We say that a proposition letter \emph{$p$ is reachable} from a vertex
$v$ in $\G$ if there is some vertex $u$ that is reachable from $v$ in
$\G$ with $L(u) = p$ or $L(u) = \neg p$. The \emph{priority} of a path
(or cycle) $p = v_1 \ldots v_n$ is defined by
\[
 \Omega(p) = \mathrm{max}\left(\{-1\} \cup \{\Omega(v_i)
\mid 1 \leq i \leq n \} \right),
\]
i.e., $\Omega(p) = -1$ if $\Omega$ is undefined on all the $v_i$.
\end{definition}


Due to the close connection between formulas and syntax graphs,
we can define an acceptance game for syntax graphs in essentially
the same way as in Definition~\ref{def:game_sem}, using that successors in the
syntax graph can be viewed as subformulas.

\begin{definition}\label{def:graph_game_sem}
 Let $\G = (V,E,L,\Omega,v_I)$ be a pointed syntax graph and
$\M=(S,\gamma,\Val,s_I)$ be a pointed game model. We define the
\emph{acceptance game} $\Acc = \Acc(\G,\M)$ as a parity game with the
game board as specified in Fig.~\ref{figure acceptance game}, initial
position $(v_I,s_I)$ and priority function $\Omega_\Acc$ such that
$\Omega_\Acc(v,s) = \Omega(v)$ if $\Omega(v)$ is defined and
$\Omega_\Acc(v,s) = 0$ otherwise.
%
If \Heloise has a winning strategy in the game $\Acc(\G,\M)$ then we say
that $\G$ \emph{accepts} $\M$. We also write $\M,s \models \G$ to mean
that \Heloise has a winning strategy in the game $\Acc(\G,\M)$ starting
from position $(v_I,s)$.

Given a pointed syntax graph $\G$ and a formula $\phi$,
we write $\G \equiv \phi$ if for all $\M$,
\Heloise has a winning strategy in $\Ev(\M,\phi)$ iff
she has one in $\Acc(\G,\M)$.
\end{definition}

\begin{figure}[!ht]
\begin{center}
\begin{tabular}{|l|c|l|}
\hline
Position $b$ & $P(b)$ & Moves $E[b]$ \\
\hline
\hline
$(v,s), L(v) = p, s \in \Val(p)$ & \Ab & $\emptyset$ \\
\hline
$(v,s), L(v) = p, s \notin \Val(p)$ & \He & $\emptyset$ \\
\hline
$(v,s), L(v) = \neg p, s \in \Val(p)$ & \He & $\emptyset$ \\
\hline
$(v,s), L(v) = \neg p, s \notin \Val(p)$ & \Ab & $\emptyset$ \\
\hline
$(v,s), L(v) = \land$ & \Ab & $\{(w_0,s),(w_1,s)\}$, where
$E(v) = \{w_0,w_1\}$ \\
\hline
$(v,s), L(v) = \lor$ & \He & $\{(w_0,s),(w_1,s)\}$, where
$E(v) = \{w_0,w_1\}$ \\
\hline
$(v,s), L(v) = \diam{g}$ & \He & $\{(v,U) \mid U \in
\nsem{g}(s) \}$ \\
\hline
$(v,U), L(v) = \diam{g}$ & \Ab & $\{(w,s) \mid s \in U, L(v) = \{w\}\}$ \\
\hline
$(v,s), L(v) = \diam{g^d}$ & \Ab & $\{(v,U) \mid U \in
\nsem{g}(s) \}$ \\
\hline
$(v,U), L(v) = \diam{g^d}$ & \He & $\{(w,s) \mid s \in U, L(v) = \{w\}\}$ \\
\hline
\end{tabular}
\end{center}
\caption{Game board of the acceptance game $\Acc(\G,\M)$}\label{figure acceptance game}
\end{figure}

A syntax graph is essentially a multi-modal version of an
alternating tree automaton (ATA) with
partial priority function as described in \cite[sec.~2.2.5]{Wilke01:ATA}.
Namely,
taking the transition graph of an ATA as defined in
\cite[sec.~2.2.4]{Wilke01:ATA} and equipping this graph with the evident
labelling function, yields a syntax graph.
Conversely, given a syntax graph one constructs for each vertex
a transition condition from its label and successors in the obvious manner.
If desired,
a partial priority function $\Omega$ can be made into a total map $\Omega'$
by defining $\Omega'(v) = \Omega(v)+2$ if $v \in V_P$ and $\Omega'(v)=0$ otherwise.
One easily adapts the notion of a run on a pointed Kripke structure from \cite{Wilke01:ATA}
to a run on a pointed game model (by dealing with modal transition conditions
as in the modal positions of Definition~\ref{def:graph_game_sem}) such that
there exists an accepting run for the ATA on $\M$ iff
\Heloise has a winning strategy in the acceptance game for the
corresponding syntax graph on $\M$.  

As described in \cite[sec.~2.2.5]{Wilke01:ATA} and in more detail in \cite[sec.~9.3.4]{Kirsten02:ATA}
ATAs can be generalised to allow complex transition conditions (i.e. arbitrary formulas)
without increasing their expressive power.
The basic idea in transforming an ATA with complex transition condition into an equivalent ATA
is to introduce new states for each node in the syntax tree of the transition conditions.

Monotone modal automata are obtained by instantiating the definition of $\Lambda$-automaton from
\cite{font:auto10} with the functor $\Mon^\AtG$ and taking $\Lambda$ to be
a suitable set of predicate liftings.
Monotone modal automata and their unguarded variants are expressively complete for the monotone (multi-modal) $\mu$-calculus. 
On the other hand,
unguarded monotone modal automata are essentially the same as
ATAs with complex transition condition (running on monotone neighbourhood models for a multi-modal signature),
hence by the above transformation,
unguarded monotone modal automata can be viewed as syntax graphs, and vice versa.

We have chosen to work with syntax graphs rather than ATAs or monotone modal automata,
since we characterise the game logic fragment mainly in terms
of the graph structure.
In the following section, we identify a class $\GLGraphs$ of syntax graphs that correspond to
game logic formulas. 
By the correspondence just outlined,
we can define game automata as those
unguarded monotone  modal automata for which the corresponding syntax graph (ATA) is in $\GLGraphs$.


\section{The Game Logic Fragment}
\label{sec:fragment}

In this section we define game logic graphs, which are a class of syntax
graphs that has the same expressivity over neighbourhood frames as
formulas in game logic. After giving the definition of game logic graphs,
we show that for each game logic formula there is a game logic
graph that accepts a pointed game model iff the formula is true at the
model and, vice versa, for every game logic graph there is a game logic formula
that is true at a pointed game model iff the game logic
graphs accepts the model.

\subsection{Game Logic Graphs}


The idea behind the definition of game logic graphs is that cycles in
the graph correspond to formulas of the form $\diam{\iter{\alpha}}\phi$
and $\diam{\diter{\alpha}}\phi$. Consider e.g. the axiom for
$\diam{\iter{\alpha}}\phi$ (in Lem.~\ref{lem:axioms}). We see that the
vertex $v$ corresponding to the disjunction in $\phi \lor
\diam{\alpha}\diam{\iter{\alpha}}\phi$ has a special role as a vertex on
the corresponding cycle. Namely, let $v_l$ and $v_r$ be the two
successors of $v$ where going to $v_l$ means \emph{leaving} the cycle (going to
subformula $\phi$) and going to $v_r$ means \emph{remaining} on the cycle
(going to subformula $\diam{\alpha}\diam{\iter{\alpha}}\phi$). We will
refer to this $v$ as the \emph{head} of the cycle corresponding to
$\diam{\iter{\alpha}}\phi$. If the cycles in the syntax graph arise from
a nesting of fixpoint formulas, and $\Omega$ is the parity function of
some formula (cf.~Def.~\ref{omega on fixpoints}), then certain
conditions will need to hold for the cycles and $\Omega$. This is made
precise in the following definition.


\begin{definition} \label{game logic graph}
Given a syntax graph $\G = (V,E,L,\Omega)$ in which $\Omega$ is
injective, we let $h := \Omega^{-1} : \range{\Omega} \to V$ denote the
inverse of $\Omega$ on its range.
We use the abbreviation $h_n := h(n)$ and call $h_n$ the \emph{head of priority $n$}.
Whenever we write $h_n$,
we presuppose that $n \in \range{\Omega}$.

A \emph{game logic graph} is a syntax graph $\G = (V,E,L,\Omega)$ in which
$\Omega$ is injective and the following conditions hold for all $n \in \range{\Omega}$:
\begin{description}
 \item[\matchingConnective] $L(h_n) = \lor$ if $n$ is odd and $L(h_n) = \land$ if $n$ is
even. \label{matching connective}
\item[\brexit] There are maps $r,l : \range{\Omega} \to V$, for which we also
use the abbreviations $r_n := r(n)$ and $l_n := l(n)$, such that $E(h_n)
= \{l_n,r_n\}$ and \label{brexit}
\begin{description}
 \item[\leavingCycles] For all simple paths $p = l_n \dots h_n$ we have that $\Omega(p)
> n$. \label{leaving cycles}
 \item[\noRemainPathToHigher] There is no simple path $h_n r_n \dots h_m$
for any $m > n$.\label{no remain path to higher}
\end{description}
\end{description}

A \emph{game logic graph with exit} is a syntax graph with exit $\G = (V,E,L,\Omega,e)$
for which $(V,E,L,\Omega)$ is a game logic graph that additionally
satisfies:
\begin{description}
\item[\exitPath]
  For all $n \in \range{\Omega}$
  and all $v \in V$ with $L(v) = e$,
  there is no simple path $h_n r_n \dots v$.
\label{exit path}
\end{description}
\end{definition}




\subsection{From Formulas to Game Logic Graphs}
\label{from formulas to graphs}

Our first result in characterising the game logic fragment of syntax
graphs shows that we can translate game logic formulas into equivalent
game logic graphs.


\def\stateThmFmlToGr{
  For every game $\alpha \in \GtDNNF$
  in which the proposition letter $e$ does not occur,
  there is a pointed syntax graph
  $\G$ with exit $e$ such that $\G \equiv \diam{\alpha} e$.
 For every game logic formula $\varphi \in \FormDNNF$ there is a pointed
syntax graph $\G$ such that $\G \equiv \varphi$.
}

\begin{theorem}\label{thm:fml_to_gr}
\stateThmFmlToGr
\end{theorem}

The proof of Theorem~\ref{thm:fml_to_gr} is by a mutual induction on the
structure of games and formulas, and is similar to the construction
of a nondeterministic finite automaton from a regular expression
\cite{houl:intr79}, that is, we define constructions on syntax graphs
that correspond to game operations and logical connectives. The
recursive procedure itself is similar to the translation of game logic
into the $\mu$-calculus \cite{Pau:phd}, with the difference that we
directly translate into syntax graphs instead of formulas of the
$\mu$-calculus.

For example, we construct
$\G_1 \comp \G_2$ where $\G_1$ and $\G_2$ are given by the induction hypothesis
by rerouting the edges that went to an exit vertex in $\G_1$ to go to the initial state
of $\G_2$.
The priority function $\Omega$ for $\G_1 \comp \G_2$
is unchanged on the $\G_2$ part, but in order to make sure
$\Omega$ is injective we shift all priority values in $\G_1$ by adding to them
a number $k$ that preserves the parity and ensures that all priorities in the $\G_1$ part
are higher than those in the $\G_2$ part. The correctness of the construction is proved by
constructing winning strategies in the evaluation game from winning strategies in the acceptance game,
and vice versa. A detailed proof is provided in \cite{dali:arxiv}.

\begin{example} \label{first example}
  Below we show the syntax graphs of some formulas.
  The initial vertex is the topmost vertex, and 
  priorities are indicated as subscripts on the vertex labels.
\begin{center}
  \begin{tabular}{ccc}
\begin{tikzpicture}[scale=0.75,transform shape]
  \Vertex[x=0,y=0,L=$\land_2$]{A}
  \Vertex[x=-1,y=-1,L=$\land$]{B}
  \Vertex[x=1,y=-1,L=$e$]{C}
  \Vertex[x=-2,y=-2,L=$\lor_1$]{D}
  \Vertex[x=0,y=-2,L=$p$]{E}
  \Vertex[x=-2,y=-3.5,L=$\diam{g}$]{F}
  \tikzstyle{LabelStyle}=[fill=white,sloped]
  \tikzstyle{EdgeStyle}=[post]
  \Edge(A)(B)
  \Edge(A)(C)
  \Edge(B)(D)
  \Edge(B)(E)
   \tikzstyle{EdgeStyle}=[bend left, post]
  \Edge(D)(F)
  \Edge(F)(D)
  \tikzstyle{EdgeStyle}=[bend left = 60, post]
  \Edge(D)(A)
\end{tikzpicture}
& \qquad\;
\begin{tikzpicture}[scale=0.75,transform shape]
  \Vertex[x=0,y=0,L=$\lor$]{A}
  \Vertex[x=-1,y=-1,L=$\lor$]{B}
  \Vertex[x=1,y=-1,L=$\land$]{C}
  \Vertex[x=-1.5,y=-2.5,L=$p$]{D}
  \Vertex[x=-0.5,y=-2.5,L=$\diam{g}$]{E}
  \Vertex[x=0.5,y=-2.5,L=$q$]{F}
  \Vertex[x=1.5,y=-2.5,L=$e$]{G}
  \Vertex[x=-0.5,y=-4,L=$e$]{H}
  \tikzstyle{LabelStyle}=[fill=white,sloped]
  \tikzstyle{EdgeStyle}=[post]
  \Edge(A)(B)
  \Edge(A)(C)
  \Edge(B)(D)
  \Edge(B)(E)
  \Edge(C)(F)
  \Edge(C)(G)
  \Edge(E)(H)
\end{tikzpicture}
& \qquad\;
\begin{tikzpicture}[scale=0.75,transform shape]
  \Vertex[x=0,y=0,L=$\land_2$]{A}
  \Vertex[x=-1,y=-1,L=$\lor$]{B}
  \Vertex[x=1,y=-1,L=$e$]{C}
  \Vertex[x=-2,y=-2,L=$\lor_1$]{D}
  \Vertex[x=0,y=-2,L=$\diam{h}$]{E}
  \Vertex[x=-3,y=-3,L=$\land$]{F}
  \Vertex[x=-4,y=-4,L=$\lnot p$]{G}
  \Vertex[x=-2,y=-4,L=$\diam{g}$]{H}
  \tikzstyle{LabelStyle}=[fill=white,sloped]
  \tikzstyle{EdgeStyle}=[post]
  \Edge(A)(B)
  \Edge(A)(C)
  \Edge(B)(D)
  \Edge(B)(E)
  \Edge(E)(A)
  \Edge(D)(F)
  \Edge(F)(G)
  \Edge(F)(H)
  \Edge(H)(D)
  \tikzstyle{EdgeStyle}=[bend left = 60, post]
  \Edge(D)(A)
\end{tikzpicture}
\\
$\varphi = \diam{\kcross{(p?;\iter{g})}}e$ &
$\varphi = \diam{(p !;g) \cup q?}e$ &
$\varphi = \diam{\diter{(\iter{((\lnot p)?;g)} \cup h)}}e$
\end{tabular}
  \end{center}
\end{example}
\def\proofThmFmlToGr{
\begin{proof}
  The proof is by a mutual induction on the structure of games and
formulas. As mentioned before, games will correspond to syntax
graphs with exit, and formulas will correspond to syntax graphs. The
base cases consist of atomic games, their duals and literals. For all
cases we describe the construction of the syntax graph and we argue that
it satisfies the conditions from Definition~\ref{game logic graph}. We
leave it to the reader to verify, using the game semantics from
Definitions \ref{def:game_sem}~and~\ref{def:graph_game_sem} that that
construction is adequate in the sense that $\G \equiv \diam{\alpha} e$
in the case of games and that $\G \equiv \varphi$ in the case for
formulas. These arguments are tedious but in most cases
straight-forward.
  
  \emph{Atomic games.}
  For $\alpha = g$ or $\alpha=g^d$ where $g \in \AtomicGames$, 
  we define the pointed syntax graph with exit
  $\G = (V,E,L,\Omega,e,v_I)$ by taking
  $V = \{v_I,w\}$ to be any two element set, 
  $E = \{(v_I,w)\}$, 
  $L(v_I) = \diam{\alpha}$ and $L(w) = e$.
  The priority function $\Omega$ can be taken to be the empty map,
  because there are no cycles in $(V,E)$.
  Hence, $\G$ is trivially a game logic graph with exit $e$.

  \emph{Literals.}
  For $\phi = p$ or $\alpha =\lnot p$ for  $p \in \Prop$,
  we define the pointed syntax graph $\G = (V,E,L,\Omega,v_I)$ by taking
  $V = \{v\}$ to be any singleton set,
  $E(v) = \emptyset$,
  $L(v) = \alpha$, $v_I = v$ and $\Omega$ as the empty map, since there are no cycles in $(V,E)$.
  Again, $\G$ is trivially a game logic graph.

  For the induction step, we define operations on syntax graphs that correspond to game constructs
  and formula constructs. 

  \emph{Demonic choice.}
  Let $\alpha = \alpha_1 \cap \alpha_2$ be such that $e$ does not occur in $\alpha$.
  By the induction hypothesis, there are 
  pointed syntax graphs with exit
  $\G_1 = (V_1,E_1,L_1,\Omega_1,e,{v_I}_1)$ and
  $\G_2 = (V_2,E_2,L_2,\Omega_2,e,{v_I}_2)$
  such that $\G_1 \equiv \diam{\alpha_1}e$ and $\G_2 \equiv \diam{\alpha_2}e$.
  We define 
  $\G_1 \cap \G_2 = (V,E,L,\Omega,e,v_I)$ by taking
  $V = \{v_I\} + V_1 + V_2$ and 
  $E = E_1 \cup E_2 \cup \{(v_I,{v_I}_1),(v_I,{v_I}_2)\}$, and
  defining the labelling by
  $L(v_I) = \land$,
  $L(v) = L_i(v)$ if $v \in V_i$ for $i=1,2$.
When defining the priority function $\Omega$ we lift all the priorities
in $\G_1$ above all those in $\G_2$ to guarantee that $\Omega$ is injective.
Thus, let $k$ be the smallest even number that is strictly
larger than the maximal priority occurring in $\G_2$. Then set
\[
 \Omega(v) = \left\{ \begin{array}{ll}
k + \Omega_1(v) & \mbox{ if } v \in V_1 \mbox{ and }
\Omega_1(v) \mbox{ is defined}, \\
\Omega_2(v) & \mbox{ if } v \in V_2 \mbox{ and }
\Omega_2(v) \mbox{ is defined}. \\
\end{array} \right.
\]
Since the cycles of $\G_1$ and $\G_2$ cannot interfere with each other,
$\G_1 \cap \G_2$ is a game logic graph with exit $e$.
%


\emph{Angelic choice.}
The syntax graph $\G_1 \cup \G_2$ is defined analogously to $\G_1 \cap
\G_2$ above, just replacing the conjunction at the initial state with a
disjunction.

\emph{Composition.}
Let $\alpha = \alpha_1 \comp \alpha_2$.
By induction hypothesis there are
pointed syntax graphs with exit
$\G_1 = (V_1,E_1,L_1,\Omega_1,e,{v_I}_1)$ and
$\G_2 = (V_2,E_2,L_2,\Omega_2,e,{v_I}_2)$
such that
$\G_1 \equiv \diam{\alpha_1}e$ and
$\G_2 \equiv \diam{\alpha_2}e$.
We define
$\G_1 \comp \G_2 = (V,E,L,\Omega,e,v_I)$ by taking
$V = V_1 + V_2$ and
\[
 E(v) = \left\{ \begin{array}{ll}
g[E_1(v)] & \mbox{ if } v \in V_1, \\
E_2(v) & \mbox{ if } v \in V_2  \\
\end{array} \right.
\qquad
\mbox{where }
 g(v) = \left\{ \begin{array}{ll}
{v_I}_2 & \mbox{ if } L(v) = e, \\
v & \mbox{ if } L(v) \neq e.
\end{array} \right.
\]
That is, we reroute all edges in $\G_1$ that
lead to the exit of $\G_1$ to the initial state of $\G_2$.
Note that all vertices that were labelled with $e$ in $\G_1$ become
  unreachable from $v_I$ in ${\G_1} \comp \G_2$.
The labelling is unchanged, i.e., $L = L_1 \cup L_2$.
%
Again, to ensure the priority map is injective,
let $k$ be the
smallest even number that is strictly larger than the maximal priority
occurring in $\G_2$, and define
\[
 \Omega(v) = \left\{ \begin{array}{ll}
k + \Omega_1(v) & \mbox{ if } v \in V_1 \mbox{ and }
\Omega_1(v) \mbox{ is defined}, \\
\Omega_2(v) & \mbox{ if } v \in V_2 \mbox{ and }
\Omega_2(v) \mbox{ is defined}. \\
\end{array} \right.
\]
Finally, we set $v_I = {v_I}_1$.
We check that $\G_1 \comp \G_2$ satisfies the 
conditions from Definition~\ref{game logic graph}.
Condition~\matchingConnective holds, since
it holds for $\G_1$ and $\G_2$.
Condition~\leavingCycles holds since
every cycle of form $h_n l_n \dots h_n$ in $\G_1 \comp \G_2$ either stays
completely $\G_1$ or completely in $\G_2$, because there are no
connections from the $\G_2$-part into the $\G_1$-part of
$\G_1 \comp \G_2$. To check Condition~\noRemainPathToHigher, note that all
priorities in the $\G_2$-part of $\G_1 \comp \G_2$ are lower than those
in the $\G_1$-part and there are no paths from the $\G_2$-part into the
$\G_1$-part. To see that Condition~\exitPath holds observe
that every path from a head in the $\G_1$-part of $\G_1 \comp \G_2$ to
the exit $e$ in $\G_1 \comp \G_2$ needs to pass through the $\G_2$-part and
hence by
the definition of $E$ gives rise to a path to an exit $e$ in $\G_1$.

\emph{Angelic tests.}
Let $\alpha = \phi ?$ for some $\phi \in \FormDNNF$
in which $e$ does not occur.
By induction hypothesis,
there is a pointed syntax graph $\G = (V,E,L,\Omega,v_I)$
with $\G \equiv \phi$.
We define a pointed syntax graph with exit
$\G? = (V',E',L',\Omega',e,v_I')$
such that $\G ? \equiv \diam{\phi ?}e$
using the axiom $\diam{\phi ?}\psi \lra \phi \land\psi$.

We define $\G?$ by adding a new initial vertex $v_I$ and an auxiliary
vertex $w'$, i.e., we take
$V' = \{v_I',w'\} + V$,
where $v_I'$ and $w'$ are distinct, and 
$E' = \{(v_I',v_I),(v_I',w')\} \cup E$.
The labelling function is defined by
$L'(v_I') = \land$,
$L'(w') = e$,
and for all $v \in V$, $L'(v) = L(v)$.
Finally, we take $\Omega' = \Omega$.
Since all cycles in $\G?$ are already cycles in $\G$,
 it follows that $\G?$ satisfies Conditions \brexit, \leavingCycles
 and \noRemainPathToHigher, since $\G$ does.
 To see that Condition~\exitPath holds observe that the new exit vertex
$w'$ is not reachable from any of the heads in the $\G$-part of $\G?$

 \emph{Demonic tests.}
 We define $\G !$ analogously to $\G?$, replacing the conjunction at $v_I'$ with a disjunction.

\emph{Angelic iteration.}
Let $\alpha = \beta^*$.
By induction hypothesis, there is a 
a pointed syntax graph with exit $\G = (V,E,L,\Omega,e,v_I)$
such that $\G \equiv \diam{\beta}e$.
We define $\G^* = (V',E',L',\Omega',e,v_I')$
using that $\diam{\alpha^*}e \lra e \lor \diam{\alpha}\diam{\alpha^*}e$.
We take
$V' = \{v_I',w'\} + V$
where $v_I'$ and $w'$ are distinct,
and define the labelling by
$L'(v_I') = \lor$,
$L'(w') = e$ and
$L'(v) = L(v)$ for all $v \in V$.
The edge relation is given by adding edges from $v_I'$ to $v_I$ and $w'$
and rerouting edges in $\G$ with an exit as destination to have destination
$v_I'$:
$E'(v_I') = \{v_I,w'\}$,
$E(w') = \emptyset$ and
for all $v \in V$: $E(v) = g[E(v)]$
where
$g(v) =  v_I'$ if $L(v) = e$,
and $g(v) = v$  if  $L(v) \neq e$.
Note that all vertices that were labelled with $e$ in $\G$ become
  unreachable from $v_I'$ in $\iter{\G}$.
To define the priority function let $n$ be the smallest odd priority
that is strictly larger than any priorities occurring in $\G$, and define
$\Omega'$ by
\[
 \Omega'(v') = \left\{ \begin{array}{ll}
n & \mbox{ if } v' = v'_I, \\
\Omega(v') & \mbox{ if } v' \in V \mbox{ and } \Omega(v') \mbox{ is
defined}.
\end{array} \right.
\]
This definition of $\G^*$ satisfies the priority condition for syntax graphs,
because all simple cycles in $\G^*$ either stay completely inside $\G$ in which
case they satisfy the priority condition because $\G$ does, or they go
out of $\G$, in which case they pass through $v'_I$ and hence contain a
vertex in the domain of $\Omega'$.
Condition~\matchingConnective is satisfied at $h_n$ because $n$ is
chosen to be odd and $L(h_n) = \lor$. For all other heads in $\G$ it is
satisfied because $\G$ is a game logic graph.

To check Condition~\brexit we let $l_n = w'$ and $r_n =
v_I$. Condition~\leavingCycles is then trivially satisfied
for $h_n$. It is satisfied for all other heads $h'_m$ in the $V$-part
because either the path stays entirely in the $\G$-part in which
case the condition is satisfied by the assumption on $\G$, or the cycles
leaves the $\G$-part in which
case it passes through $h_n$ with priority $n$.

Condition~\noRemainPathToHigher is satisfied by the head $h_n$
because by the choice of $n$ there are no heads of higher priority than
$n$. To see that Condition~\noRemainPathToHigher is satisfied
by all other heads $h_m$, with $m < n$ in the $\G$-part assume for a
contradiction that there is a simple path $p = h_m r_m \dots h_k$ for some
$k > m$. If $p$ stays entirely in the $\G$-part of $\G^*$ then we obtain
a contradiction with the assumption that $\G$ is a game logic graph. So
we only need to consider the case where $p$ at one point leaves the
$\G$-part. But by inspecting the definition of $E$ one finds that this
can only happen if in $\G$ there is a corresponding path $q = h_m r_m \dots
v$ where $L(v) = e$, which with the exception of the end point is an initial
segment of the simple path $p$. Hence, $q$ is almost a simple path with
the possible exception that there might be a contraction with the end
point $v$. Because $L(h_m) \neq e$, since $h_m$ is a head, we know that
such a contraction again leads to a path of the form $h_m r_m \dots v$.
As we can repeat this contractions until we obtain a simple path we can
then assume that $q = h_m r_m \dots v$ is actually a simple path in $\G$
where $L(v) = e$. This contradicts Condition~\exitPath for the
game logic graph with exit $\G$.

Last, we check that Condition~\exitPath is satisfied by $\iter{\G}$
with exit $e$. This is trivially the case for the head $h_n = v_I'$. For
any other head we might reason similarly to the last part of the
argument in the previous section. If there was a simple path
$r_m \dots w'$ then this path must pass through $v_I'$
and hence there would be a path in $\G$ of the form
$h_m r_m \dots v$ in $\G$ with $L(v) = e$,
which contradicts the Condition~\exitPath for $\G$.

Finally, let us make a remark about how to show that $\iter{\G} \equiv
\diam{\iter{\beta}}e$. The argument uses the fixpoint game from
Definition~\ref{fixpoint game} and Lemma~\ref{lem:fixgames}. It is
similar to the reasoning about $\iter{\lambda}$ in the proof of
Theorem~\ref{t:sftofm} below.

\emph{Demonic iteration.}
We define $\G^\cross$ similarly to $\iter{\G}$,
but we label the initial node $v_I'$ with $\land$ instead of $\lor$,
and we take $n$ to be the least even priority that is strictly larger
than any priority in $\G$.

\emph{Conjunction and disjunction.}
In the case $\phi = \phi_1 \land \phi_2$ or $\phi = \phi_1 \lor \phi_2$, 
we define $\G_1 \land \G_2$ and $\G_1 \lor \G_2$
of two pointed syntax graphs $\G_1$ and $\G_2$
basically in the same way as $\G_1 \cap \G_2$ and  $\G_1 \cup \G_2$,
except that we do not need to handle the exits.

\emph{Modal operators.}
In the case $\phi = \diam{\alpha}\psi$,
we obtain by induction hypothesis 
a pointed syntax graph  $\G_1$ with exit $e$ and a pointed syntax graph $\G_2$
such that $\G_1 \equiv \diam{\alpha}e$ and $\G_2 \equiv\psi$.
We define
the pointed syntax graph $\diam{\G_1} \G_2$ 
similarly to the composition $\G_1 \comp \G_2$,
but since there is no exit in $\G_2$
there is also no exit in $\diam{\G_1} \G_2$. \qed
\end{proof}
}
\ifWithProofs
\proofThmFmlToGr
\fi

\newcommand{\F}{\mathcal{F}}
\newcommand{\transition}{\sigma}

\renewcommand{\S}{\DONTUSES}

\subsection{From Game Logic Graphs to Formulas}
\label{graphs to formulas}

\newcommand{\bbG}{\mathbb{G}}
\newcommand{\bbS}{\mathbb{S}}
\newcommand{\yvl}{\underline{\lambda^{*}}}
\newcommand{\EG}{\mathcal{E}}
\newcommand{\GG}{\mathcal{F}}
\newcommand{\sat}{\Vdash}
\newcommand{\eloi}{\exists}
\newcommand{\Dom}{\mathrm{dom}}

We now show how to transform game logic graphs into equivalent game
logic formulas.

\def\stateThmGrToFml{
 For every pointed \red{game logic} graph with exit $\G = (V,E,L,\Omega,e,v_I)$
there is a game term $\delta \in \GT$, not containing $e$ and only
containing propositional letters that are reachable from $v_I$, such that $\G \equiv
\diam{\delta}{e}$.
}

\begin{theorem}
\label{t:sftofm}
\stateThmGrToFml
\end{theorem}

The proof of Theorem~\ref{t:sftofm} is by induction on the number of
heads in the \red{game logic} graph. In the base case there are no heads which
implies that there are no cycles in the graph, which makes it easy to
recursively decompose the graph into a game term. In the inductive step
we use a construction that removes some of the edges at the head with
the highest priority and thus cutting all cycles that pass through the
highest priority head. This allows us to remove the priority from this
head and obtain a simpler \red{game logic} graph to which we can apply the
induction hypothesis. A detailed proof is provided in \cite{dali:arxiv}.

\def\proofThmGrToFml{
\begin{proof}
We prove the theorem by induction on the size of the domain of $\Omega$.

In case $\Dom(\Omega) = \emptyset$, this means that $\mathbb{G}$ has no cycles, 
and so we can do a straightforward subinduction on the edge relation (or more
precisely, on the well-order consisting of the inverse of the transitive closure
of the edge relation). 
That is, we can define, for every vertex $v \in V$ a game $\delta_{v}$ as 
follows:

   \[\begin{array}{lll}
   \delta_{v} \isdef
     \left\{\begin{array}{ll}
        (L(v)?)\comp (L(v)!)
            & \text{ if $L(v)$ is a literal distinct from $e$}
     \\ (p \lor \neg p)?
            & \text{ if $L(v) = e$ and $p \neq e$}
     \\ \delta_{v_1} \cap \delta_{v_2}
            & \text{ if $L(v) = \land$ and $E(v) = \{ v_1, v_2\}$}
     \\ \delta_{v_1} \cup \delta_{v_2}
            & \text{ if $L(v) = \lor$ and $E(v) = \{ v_1, v_2\}$}
     \\ L(v); \delta_{u}
            & \text{ if $L(v) \in \{ g, g^d \mid g \in \AtomicGames \}$ and 
	         $E(v) = \{ u \}$}
     \end{array}\right.
   \end{array}\]
It is routine to prove that this definition is correct, i.e., that
$\mathbb{G}@v \equiv \diam{\delta_v}e$ for all $v \in V$. From this it follows that $\bbG
\equiv
\diam{\delta_{v_I}} e$

In case $\Dom(\Omega) \neq \emptyset$ we may by assumption consider the
head $h$ with highest priority in $\Dom(\Omega)$; let $l$ and $r$ be the
successors of $h$ (i.e., we omit subscripts).

Let us first note for further reference that:
\begin{equation} \label{h not from l}
 h \mbox{ is not reachable from } l \mbox{ in } \bbG . 
\end{equation}
 To see that $h = h_n$ is not reachable from $l = l_n$ in $\G$ assume for
a contradiction that there is a path $p = l_n \dots h_n$ in $\G$. We can
contract duplications of vertices on $p$ until we have a simple path $p'
= l_n \dots h_n$ in $\G$. By Condition~\leavingCycles it follows that
$\Omega(p') > n$, which is impossible because $n$ was chose to be the
maximal priority occurring in $\G$.

We define the two pointed syntax graphs with exit $\G'$ and $\G''$:
Let $\yvl$ be a fresh atomic game, and let $e'$ be a fresh variable.

\begin{itemize}
\item
$\bbG' \isdef (V,E',L',\Omega',e,v_I)$, where 
\[\begin{array}{lll}
   E' & \isdef & E \setminus \{ (h,r) \} 
\\ L'(u) & \isdef &
    \left\{\begin{array}{ll}
       L(u) & \text{if } u \neq h
    \\ \yvl & \text{if } u = h
    \end{array}\right.
\\ \Omega'(u) & \isdef  &
    \left\{\begin{array}{ll}
       \Omega(u)  & \text{if } u \neq h
    \\ {\uparrow} & \text{if } u = h
    \end{array}\right.
\end{array}\]
In words, we obtain $\bbG'$ from $\bbG$ by dropping the edge from $h$ to $r$,
relabelling $h$ with the dummy atomic game $\yvl$, and removing it from the 
domain of $\Omega$.

Observe that every path that exists in $\G'$ also exists in $\G$.

We need to argue that $\G'$ is a game logic graph with exit.

It is not immediately obvious that $\G'$ satisfies the priority
condition on syntax graphs because we removed the priority from the
state $h$. This is not a problem, however, because if there was any
cycle in $\G'$ that passes through $h$ it would by the definition of
$E'$ need to continue to $l$ and hence lead to a contradiction with
\eqref{h not from l}.

It satisfies Condition~\matchingConnective because all it the
remaining heads in $\G'$ are also heads in $\G$ and their label did not
change.

To see that $\bbG'$ satisfies Condition~\leavingCycles take any
simple path $p = l_m \dots h_m$ in $\G'$. Because $h = h_n$, where $n$
is maximal priority in $\G$, we have that $m < n$. This path $p$ also
exists $\G$. We distinguish cases depending on whether $h_n$ lies on
$p$. If $h_n$ does not lie on $p$ then clearly $\Omega'(p) = \Omega(p)$
and $\Omega(p) > m$ because $\G'$ satisfies Condition~\leavingCycles. We can show that it is not possible that $h$ lies on $p$.
Assume this was the case. Then $h$ is followed by $l = l_n$ on $p$, because $p$
is also a path in $\G'$ and the only connection out of $h$ in $\G'$ is
via $l$. So we have a path of the form $p = l_m \dots h_n l_n \dots
h_m$. Because there is an edge from $h_m$ to $l_m$ we can rotate this
path until it is of the form $p' = l_n \dots h_m l_n \dots h_n$.
By
Condition~\leavingCycles it follows that $\Omega(p') > n$, which
is not possible, because $n$ is the maximal priority in $\G$.

Condition~\noRemainPathToHigher holds in $\G'$ because any path of
form $h_m r_m \dots h_k$ with $k > m$ in $\G'$ would also exist and
violate Condition~\noRemainPathToHigher in $\G$.

Similarly, Condition~\exitPath holds in $\G'$ because any
violating path would also do so in $\G$.

\item
$\bbG'' \isdef (V,E'',L'',\Omega'',e',r)$, where $e'$ is a fresh
propositional letter not occurring in $\G$ and
\[\begin{array}{lll}
   E'' & \isdef & E \setminus \{ (h,r), (h,l) \} 
\\ L''(u) & \isdef &
    \left\{\begin{array}{ll}
       L(u) & \text{if } u \neq h
    \\ e'    & \text{if } u = h
    \end{array}\right.
\\ \Omega''(u) & \isdef  &
    \left\{\begin{array}{ll}
       \Omega(u)  & \text{if } u \neq h
    \\ {\uparrow} & \text{if } u = h
    \end{array}\right.
\end{array}\]
In words, we obtain $\bbG''$ from $\bbG$ by initialising it at $r$, dropping the
edges from $h$ to $l$ and $r$, relabelling $h$ with a new exit variable $e'$, 
and removing $h$ from the domain of $\Omega$.

Observe that every path that exists in $\G''$ also exists in $\G$.

We need to argue that $\G''$ is a game logic graph with exit.
It satisfies Condition~\matchingConnective because all it's heads
are also heads in $\G$ and their label did not change.

Preservation of Conditions \matchingConnective~and~\ref{no
remain path to higher} from $\G$ to $\G''$ is equally trivial as in the
case of $\G'$ discussed above.

To see that Condition~\leavingCycles holds in $\G''$ consider any
path $p = l_m \dots h_m$ for some $m < n$. Note that $h = h_n$ can not lie
on $p$. It could only be its last vertex because $h$ has no successors
in $\G''$ but then $h_n = h_m$ which is impossible because $n < n$ and
the function $h$ is injective. Hence $\Omega'(p) = \Omega(p)$ and
$\Omega(p) > m$ because $\G$ satisfies Condition~\leavingCycles.

To see that Condition~\exitPath holds in $\G''$ assume for a
contradiction that there was a path $p = h_m \dots v$ with $L(v) = e$
in $\G$. Because $h_m$ is a head in $\G$ we have that $m < n$. Because
$h = h_n$ is the only vertex in $\G$ that has the propositional letter $e$
in $\G''$ it follows that $p$ is of the form $h_m \dots h_n$ in $\G$.
This is a contradiction to Condition~\leavingCycles for $\G$.
\end{itemize}

\noindent
We make one more observations about our constructions that we need later:
\begin{equation} \label{e not from r}
 e \mbox{ is not reachable from } r \mbox{ in } \G''.
\end{equation}
\noindent
To check that $e$ is not reachable from $r$ in $\G''$ assume for a
contradiction that there is a path $p = r_n \dots u$ in $\G''$ from
$r_n$ to some vertex $u$ that is labelled with $e$. This path $p$ does
not contain $h_n$ because $h_n$ has no successors in $\G''$ and $h_n$ is
distinct from the last vertex $u$ on $p$ because $h_n$ it is labelled
with $e'$ in $\G''$ which was fresh and hence distinct from the letter
$e$ that labels $u$. The path $p = r_n \dots u$ then also exists in $\G$
where it extends to a path of the form $h_n p = h_n r_n \dots u$. This
path can be contracted to a simple path that is still of this form
because no contraction is possible at the initial vertex $h_n$ since
$h_n$ does not lie on $p$. Thus we obtain a contradiction with
Condition~\exitPath for the game logic graph $\G$ with exit $e$.

\noindent
The induction hypothesis applies to both $\bbG'$ and $\bbG''$, so that:
\begin{itemize}
\item There is an $\AtomicGames \cup \{ \yvl \}$-game $\delta_{0}$ not
containing $e$ or $e'$ such that $\bbG' \equiv \diam{\delta_{0}} e$.
\item There is a $\AtomicGames$-game $\lambda$ not involving 
   $e$ or $e'$ such that $\bbG'' \equiv \diam{\lambda} e'$
\end{itemize}
To see that $\delta_0$ does not contain $e'$ observe that $e'$ is only
used in the definition of $\G''$, where it is assumed to be fresh.

To see that $\lambda$ does not contain $e$ note that the induction
hypothesis guarantees that all propositional letters that occur in
$\delta_0$ are reachable in $\G''$, but by \eqref{e not from r} we have
that $e$ is not reachable from $r$, which is the initial vertex of
$\G''$.

Now define
\[
\delta \isdef \delta_{0}[\lambda^{*}/\yvl],
\]
that is, everywhere in $\delta_{0}$ we substitute the actual game $\lambda^{*}$ 
for the formal atomic game $\yvl$.
It will be our purpose to prove that 
\begin{equation}
\label{eq:yvmain}
\bbG \equiv \diam{\delta}e.
\end{equation}

\noindent
Before we set up our proof, we need some auxiliary definitions.
First, a $\AtomicGames$-model is a game model over the set $\AtomicGames$
of atomic games.
\begin{itemize}
\item 
Given a $\AtomicGames$-model $\bbS = (S,\gamma, \Val)$, we define $\bbS^{\bullet}$ 
as the $\AtomicGames \cup \{ \yvl \}$-model $(S,\gamma^{\bullet}, \Val)$, where 
$\gamma^{\bullet}(g) \isdef \gamma(g)$ if $g \neq \yvl$ and 
$\gamma^{\bullet}(\yvl) \isdef \gamma_{\lambda^{*}}$, that is, 
$\gamma^{\bullet}(\yvl)$ is the actual interpretation of $\lambda^{*}$ by 
$\gamma$.
\end{itemize}

\noindent
The following claim then should be obvious (its proof is a standard inductive
substitution argument).

\begin{yclaim}
\label{cl:1}
For any $\AtomicGames$-model $\bbS = (S,\gamma, \Val)$ we have that 
\[
\gamma^{\bullet}(\delta_{0}) = \gamma(\delta).
\]
\end{yclaim}

\noindent
The key observation in our proof is the following claim.

\begin{yclaim}
\label{cl:key}
For any $\AtomicGames$-model $\bbS = (S,\gamma, \Val)$ and any state $s_{I} \in S$ 
we have that 
\begin{equation*}
\bbS,s_{I} \sat \bbG \iff \bbS^{\bullet},s_{I} \sat \bbG'.
\end{equation*}
\end{yclaim}

To see how \eqref{eq:yvmain} follows from this, observe that for any
$\AtomicGames$-model $\bbS = (S,\gamma, \Val)$ and any state $s_{I} \in S$ we have
\begin{align*}
\bbS,s_{I} \sat \bbG & \iff \bbS^{\bullet},s_{I} \sat \bbG'
   & \text{(Claim~\ref{cl:key})}
\\ & \iff \bbS^{\bullet},s_{I} \sat \diam{\delta_{0}}e
   & \text{(induction hypothesis on $\bbG'$)}
\\ & \iff \bbS,s_{I} \sat \diam{\delta}e
   & \text{(Claim~\ref{cl:1})}
\end{align*}

Hence it suffices to prove the key claim.

\begin{proofof}{Claim~\ref{cl:key}}
We only consider the direction from left to right, the proof for the opposite
direction being simpler/similar.
Let $f$ be a positional winning strategy for \Heloise in the evaluation game $\EG
\isdef \EG(\bbG,\bbS^{\bullet})$, and suppose that $(v_I,s_{I})$ is a winning 
position in this game.
We need to supply \Heloise with a winning strategy in the game $\EG' \isdef 
\EG(\bbG',\bbS)$, starting at position $(v_I,s_{I})$.
Observe that the game boards of these two games are almost the same: the only
difference lies in positions of the form $(h,s)$, since $L(h) = \lor \neq \yvl 
= L'(h)$. 
For this reason, the strategy $f$ can be used in $\EG'$ as well, for all 
positions $(v,s)$ where $v \neq s$.

\Heloise's strategy in position $(v_I,s)$ will be as follows:
\begin{enumerate}
\item 
She starts with playing the strategy $f$.
It is easy to see that this guarantees her to win any match that does not pass
through a position of the form $(h,s)$.
\item
If, on the other hand, at some stage a position of the form $(h,s)$ is reached, 
she continues as follows.
Let $U \subseteq S$ be the set of states $t$ such that $(h,t) \in \Win_{\eloi}
(\EG)$ and at this position $(h,t)$ \Heloise's strategy $f$ picks the left
successor of $h$:
\[
U \isdef \{ t \in S \mid (h,t) \in \Win_{\eloi}(\EG) \text{ and } f(h,t) = (l,t) \}. 
\]
We claim that $U$ is a legitimate move for \Heloise in $\EG'$ at
position $(h,s)$:
\begin{equation}
\label{eq:yv1}
U \in \gamma^{\bullet}(\yvl)(s),
\end{equation}
and let \Heloise play this move indeed.
\item
Suppose that \Abelard responds to this move by picking $t \in U$, so that $(l,t)$
is the next position in the $\EG'$-match.
Then by definition of $U$ we have that $(h,t) \in \Win_{\eloi}(\EG)$ and 
\Heloise's winning strategy $f$ in $\EG$ picks the position $(l,t)$ at $(h,t)$.
This means in particular that $(l,t)$ is a winning position for \Heloise in $\EG$.

Since $h$ is the only vertex at which $\G$ and $\G'$ differ, but from
\eqref{h not from l} we know that $h$ is not reachable from $l$ in $\G$,
it follows that the the syntax graphs $\bbG@l$ and $\bbG'@l$ are
identical. This means that from this moment on, \Heloise can resume
playing her winning $\EG$-strategy $f$ in $\EG'$ again, and it is easy
to see that any resulting full $\EG'$-match is a win for \Heloise.
\end{enumerate}

It is left to prove \eqref{eq:yv1}. 
By definition of $\gamma^{\bullet}$ this means that we need to show that 
$U \in \gamma_{\lambda^{*}}(s)$, and by Lemma~\ref{lem:fixgames}
it suffices to show that $s \in \Win_{\eloi}(\GG)$, where $\GG \isdef 
\game(\lambda^{*},\bbS,U)$ is the game defined in
Definition~\ref{fixpoint game}.
In other words, we have to supply \Heloise with a winning strategy for
position $s$ in $\GG$.

To define this strategy $\overline{f}$, take an arbitrary point $t \in S$, and 
make the following case distinction:
\begin{itemize}
\item 
If $(h,t) \not\in \Win_{\eloi}(\EG)$, \Heloise makes a random move (we will make
sure that this situation never occurs).
\item
If $(h,t) \in \Win_{\eloi}(\EG)$ and $t \in U$, $\overline{f}$ picks the empty 
set as a move for \Heloise.
Clearly this is legitimate and it causes an immediate win for \Heloise in
$\GG$.
\item
If $(h,t) \in \Win_{\eloi}(\EG)$ but $t \not\in U$, then by definition of $U$ 
\Heloise's winning strategy $f$ picks the right successor $r$ of $h$ at the 
position $(h,t)$ in $\EG$.
This means that $(r,t) \in \Win_{\eloi}(\EG)$.
Define $W_{t}$ to be the set of points $u \in S$ for which there is a partial
$f$-guided $\EG$-match $\Pi = (v_{0},s_{0}) \cdots (v_{n},s_{n})$ such that 
$(v_{0},s_{0}) = (r,t)$, $(v_{n},s_{n}) = (h,u)$ and $v_{i} \neq h$ for all 
$0<i<n$.
It should then be obvious that 
\begin{equation}
\label{eq:yv2}
W_{t} \subseteq \{ u \in S \mid (h,u) \in \Win_{\eloi}(\EG) \}.
\end{equation}
We define $\overline{f}(t) \isdef W_{t}$, that is we let $W_{t}$ be \Heloise's 
move at position $t$ in $\GG$, and first show its legitimacy:
\begin{equation}
\label{eq:yv3}
W_{t} \in \gamma_{\lambda}(t).
\end{equation}
In order to prove \eqref{eq:yv2} we turn to the evaluation game $\EG'' 
\isdef \EG(\bbG'',\bbS^{\circ})$, where $\bbS^{\circ} \isdef (S,\gamma,
\Val[e' \mapsto W_{t}])$. Here we use $\Val[e' \mapsto W_t]$ to denote the
valuation that is just like $\Val$ but maps the propositional letter $e'$
to the set $W_t \subseteq S$.
It follows from the inductive hypothesis on $\bbG''$ that 
\[
(r,t) \in \Win_{\eloi}(\EG'') \iff \bbS^{\circ},t \sat \diam{\lambda}e',
\]
so that by definition of $\bbS^{\circ}$ we find that 
\[
(r,t) \in \Win_{\eloi}(\EG'') \iff W_{t} \in \gamma_{\lambda}(t).
\]
Hence, in order to prove \eqref{eq:yv3} it suffices to show that 
\begin{equation}
\label{eq:yv4}
(r,t) \in \Win_{\eloi}(\EG''),
\end{equation}
that is, we have to supply \Heloise with a winning strategy in the 
game $\EG''$ at position $(r,t)$.
However, given the similarities between $\bbG$ and $\bbG''$ and between $\bbS$ 
and $\bbS^{\circ}$, she can simply use her $\EG$-strategy $f$ once more.
Consider a full $\EG''$-match $\Pi$ where she plays like this.
If $\Pi$ does not pass through a position of the form $(h,u)$ then this match
is also an $\EG$-match, and it is easy to see that \Heloise wins.
On the other hand, if $\Pi$ does pass through such a position $(h,u)$, this can 
only mean that in $\EG''$ this is the final position of $\Pi$.
It then follows by definition of $W_{t}$ that $u \in W_{t}$, so that $u \in 
\Val[e' \mapsto W_{f}](e')$, and again \Heloise is the winner of the $\EG''$-match 
$\Pi$.
This proves \eqref{eq:yv4}.
\end{itemize}

It remains to prove that this strategy $\overline{f}$ is winning for
\Heloise at position $s$ in $\GG$. So consider a full
$\overline{f}$-guided match $\Sigma$ in $\GG$ starting at $s$. It
follows from \eqref{eq:yv2} that all positions $t$ reached in this match
are such that $(h,t) \in \Win_{\eloi}(\EG)$, so that we can make the
following case distinction.

If some position $t$ of $\Sigma$ belongs to $U$, then $\overline{f}$ picks the 
empty set and \Heloise wins immediately.
If, on the other hand, all positions of $\Sigma$ are such that $(h,t) \in 
\Win_{\eloi}(\EG)$ but $t \not\in U$, then we only have to worry that $\Sigma$ 
might be infinite (since \Heloise has a legitimate move $W_{t}$ for all $t$ on 
$\Sigma$, $\Sigma$ being finite means that \Abelard got stuck).
So assume for contradiction that $\Sigma$ is of the form $\Sigma = 
t_{0}W_{0}t_{1}W_{1}\cdots$ with $t_{0} = s$.
Given the definition of $\overline{f}$ it is easy to see that for every $n \in
\omega$ there is a partial $f$-guided $\EG$-match $\Sigma_{n} = (h,t_{n}) 
\cdots (h,t_{n+1})$.
But then we could glue these partial matches together, forming one infinite
$\EG$-match $\Sigma' = (h,t_{0}) \cdots (h,t_{1}) \cdots$.
This provides the desired contradiction, since this $\Sigma'$ would be a loss 
for \Heloise since it passes $h$ infinitely often, whereas at the same time it is
guided by \Heloise's winning strategy $f$.
In other words, $\overline{f}$ is a winning strategy for \Heloise in the
game $\GG$ starting from $s$.
\end{proofof}

This finishes the proof of Theorem~\ref{t:sftofm}. \qed
\end{proof}
}
\ifWithProofs
\proofThmGrToFml
\fi

Because any propositional letter $e$ that does not occur in $\G$ can be
added as an exit to a game logic graph $\G$ we obtain the following
corollary from Theorem~\ref{t:sftofm}:
\begin{corollary}
 For every pointed game logic syntax graph $\G$ there is a formula
$\varphi \in \Form$ such that $\G \equiv \varphi$.
\end{corollary}

\begin{example}
 We apply the construction from Theorem~\ref{t:sftofm} to the
 graph on the left in Example~\ref{first example}.
 The heads $h_1$ and $h_2$ are
 the disjunction with priority $1$
 and the conjunction with priority $2$, respectively.
 We start the decomposition at $h_2$.
 We then first obtain a game $\delta_2 =
\underline{\lambda_2^{\cross}}$, where $\underline{\lambda_2^{\cross}}$
is a dummy game term that is a place holder for the game through the
left child of $h_2$, that describes how to reach the exit from the
initial state without iterating at $h_2$. We also apply the
induction hypothesis to obtain a new game $\lambda_2$ that describes one
iteration from the left node to $h_2$, which we replace by a fresh
exit $e'$. In this inductive step we then need to cut $h_1$.
At $h_1$ we have $\delta_1 = \underline{\lambda_1^*}
\cap p? \comp p!$ and $\lambda_1 = \diam{g}$. We then obtain $\lambda_2$
by substituting $\underline{\lambda_1^*}$ in $\delta_1$ with
$\lambda_1^*$ and thus obtain $\lambda_1 = g^* \cap (p? \comp p!)$.
Substituting $\lambda_1^\cross$ for $\underline{\lambda_1^\cross}$ in
$\delta_2$ yields the overall game $(g^* \cap (p? \comp p!))^\cross$.
Hence the game graph is equivalent to the formula $\diam{(g^* \cap (p?
\comp p!))^\cross} e$.
\end{example}

\section{Conclusion}

We have provided a semantics for game logic in terms of parity games. This was the key to obtain our main technical result, the characterisation of game logic graphs, i.e., a class of parity automata that correspond to game logic formulas. 

These automata open several avenues for future research: Firstly, we would like to study normal forms in game logic. In the $\mu$-calculus, automata are the key to obtain the (semi-)disjunctive normal forms of formulas which can be used to prove further results, e.g., completeness, interpolation and the characterisation of the expressivity of the logic~\cite{jani:auto95,Wal95,agos:logi00}. Our  experience suggests that a similar normal form for game logic is out of reach, but 
a careful analysis of the cycle structure  of game logic graphs might yield useful insights
concerning the structure of game logic formulas. As a first step in this direction we are currently investigating how to obtain guarded game logic graphs and, consequently, a definition of guarded game logic formulas.

Furthermore, game logic constitutes a very general dynamic logic that makes very few assumptions on the algebraic properties of the modal operators. Therefore we believe that our game logic automata have the potential to help us understand a wider class of automata for families of dynamic logics such as coalgebraic dynamic logics~\cite{HK:FICS15} or many-valued dynamic logics as described in~\cite{manema:anex16} or for a combination of these frameworks.


\bibliographystyle{splncs03}
\bibliography{game-logic,extra}


\ifWithProofs
    {}
\else
\newpage
\begin{appendix}
\appendix

\section{Proofs from Section 2}


\begin{lemma_for}{lem:MF-congruence}{\stateLemMFcongruence}
\end{lemma_for}
\proofLemMFcongruence

\begin{lemma_for}{lem:congruence}{\stateLemCongruence}
\end{lemma_for}
\proofLemCongruence

\begin{lemma_for}{lem:DNNF}{\stateLemDNNF}
\end{lemma_for}
\proofLemDNNF

\section{Proofs from Section 3}
As a preparation for proving Lemma~\ref{lem:fromEtoG} we need to introduce some terminology.
\begin{definition}\label{def:suc}
	Let $\varphi \in \Form$, let $\M=(S,\gamma,\Val)$ be a game model and consider
	the game $\Ev = \Ev(\M,\varphi)$. Furthermore we let $f_\E$ denote a strategy for \Heloise.
	For $(s,\diam{\alpha}\psi)$
	we denote by $\Suc_{f_\E}(s,\diam{\alpha}\psi) \subseteq S$ the collection of all states
	$s' \in S$ such that there exists a (possibly partial) $\Ev$-play that follows $f_\E$, that is of the form $(s,\diam{\alpha}\psi) \dots (s',\psi)$ and no position in between $(s,\diam{\alpha}\psi)$ and $(s',\psi)$ is of the form
	$(s'',\psi)$.
\end{definition}

\begin{remark}\label{rem:suc}
	We spell out the definition of $\Suc$ for the test cases, as these cases tend to cause unnecessary confusion.		For $\alpha= \xi?$ we have $\Suc_{f_\E}(s,\diam{\alpha}\psi) = \{s\}$.
	For $\alpha= \bang{\xi}$ we have $\Suc_{f_\E}(s,\diam{\alpha}\psi) = \emptyset$ if $f_\E$ requires 
	to move from $(s,\diam{\alpha}\psi)$ to $(s,\xi)$ and we have $\Suc_{f_\E}(s,\diam{\alpha}\psi) = \{ s \}$
	if $f_\E$ requires 
	to move from $(s,\diam{\alpha}\psi)$ to $(s,\psi)$. 
\end{remark}

\begin{lemma_for}{lem:fromEtoG}{\stateLemFromEtoG}
\end{lemma_for}
\proofLemFromEtoG

\begin{lemma_for}{lem:reachpsi}{\stateLemReachPsi}
\end{lemma_for}
\proofLemReachPsi

\begin{proposition_for}{prop:eval-game}{\statePropEvalGame}
\end{proposition_for}
\proofPropEvalGame


\section{Proofs from Section 5}


\begin{theorem_for}{thm:fml_to_gr}{\stateThmFmlToGr}
\end{theorem_for}
\proofThmFmlToGr

\begin{theorem_for}{t:sftofm}{\stateThmGrToFml}
\end{theorem_for}
\proofThmGrToFml

\end{appendix}
\fi

\end{document}